\documentclass[journal,onecolumn,draft]{IEEEtran}
\usepackage{cite}
\usepackage{float}
\usepackage{amssymb}
\usepackage{graphicx}
\usepackage{epstopdf}
\usepackage{amsmath}

\usepackage{bm}
\usepackage{color}
\usepackage{lineno,hyperref}

\ifCLASSINFOpdf
\else
\fi
\begin{document}
%
\title{Adaptive Fuzzy Control for Fractional-Order Interconnected Systems with Unknown\\ Control Directions}
%
%
%

\author{Bingyun~Liang,
        Shiqi~Zheng, and
        Choon Ki~Ahn,~\IEEEmembership{Senior Member,~IEEE}
\thanks{The work was supported  by the National Natural Science Foundation of China under Grant No. 61703376. \emph{Corresponding author: Shiqi Zheng and Choon Ki Ahn.}}
\thanks{B. Liang and S. Zheng  are with the School of Automation, China University of Geosciences, Wuhan 430074, China, and also with the Hubei Key Laboratory of Advanced Control and Intelligent Automation for Complex Systems, Wuhan 430074, China (e-mails: zhengshiqi1000@foxmail.com)}
\thanks{C. K. Ahn is with the School of Electrical Engineering, Korea University,
	Seoul 136-701, South Korea (e-mail: hironaka@korea.ac.kr).}
}

%
%

\markboth{......}
{LIANG \MakeLowercase{\textit{et al.}}: Adaptive Fuzzy Control for Fractional Order Interconnected Systems with Unknown Control Directions}
%



\maketitle

\begin{abstract}
This paper concentrates on the study of the decentralized  fuzzy control method for a class of fractional-order interconnected systems with unknown control directions. To overcome the difficulties caused by the multiple unknown  control directions in  fractional-order systems, a novel fractional-order Nussbaum function technique is proposed. This technique is much more general than those of existing works since it not only handles single/multiple unknown control directions but is also suitable for fractional/integer-order single/interconnected systems. Based on this  technique, a new decentralized adaptive control method is proposed for fractional-order interconnected systems. Smooth functions are introduced to compensate for  unknown interactions among subsystems adaptively. Furthermore,  fuzzy logic systems are utilized to approximate  unknown nonlinearities. It is proven that the designed controller can guarantee the boundedness of all signals in interconnected systems and the convergence of tracking errors. Two  examples are given to show the validity of the proposed  method.
\end{abstract}


%
\IEEEpeerreviewmaketitle

\section{Introduction}
%
%
%
%

Interconnected  systems are common in engineering applications, such as power systems, chemical processes, computer networks, and aerospace systems \cite{a0}. The decentralized control method, which relies on the local information of each subsystem, provides an effective way of handling interconnected systems because of its low complexity feature. Therefore,  decentralized control for interconnected systems has drawn much attention in the past few years.

Since the 1990s, the backstepping design technique has been widely employed for nonlinear systems \cite{a1}. \cite{a2} first proposed the decentralized control method with the backstepping technique for interconnected nonlinear systems. Afterward, many studies have reported stabilization and tracking control problems for interconnected systems. For example, with the consideration of unknown interactions,  decentralized adaptive controllers have been proposed for various kinds of interconnected systems in \cite{a8,a5,a6,interaction}. Meanwhile, in \cite{aFu,aF1,aF2,aF3,aF4,aF5}, decentralized adaptive controllers have been studied using a universal approximation method ({i.e.,} utilizing fuzzy logic systems (FLSs) to approximate interactions).

The control problem without  \emph{a priori} knowledge of the control direction is undergoing active research in the control community. The problem was first solved by adopting the Nussbaum function in \cite{u1}. Benefiting from pioneering work and the Nussbaum function, numerous control strategies have been presented for nonlinear systems with unknown control directions (e.g., \cite{u2,u5}). In \cite{u4},  unknown time-varying control gains in nonlinear systems were handled with a new Nussbaum function technique.
However, the above results were related to a single Nussbaum function technique for single-input single-output systems. The control of interconnected systems with multiple unknown control directions is more difficult because   multiple Nussbaum functions should be adopted. Based on the backstepping method, a few results on the multiple Nussbaum functions technique have been reported. For instance, \cite{u7,u8,u9,u10} have proposed new multiple Nussbaum functions  to deal with unknown time-varying coefficients.

In the last few decades, fractional-order calculus has attracted considerable attention from the control field because fractional-order differential equations can describe many physical phenomena concisely and precisely, such as servo motor systems, viscoelastic structures, and heat conduction \cite{f1}. Meanwhile, in contrast to integer-order controllers, fractional-order ones have more design freedom and robust ability \cite{fe2}. More recently, various interesting results on stability analysis and control schemes for fractional-order nonlinear systems have been reported (see \cite{f0,fe1,fe3,fe4} for examples). Moreover, in \cite{f2,f3}, researchers have taken the lead in extending the backstepping control scheme to fractional-order nonlinear systems.  \cite{f4} presented a backstepping control strategy for nonlinear fractional-order systems with the direct Lyapunov method \cite{f7,f8,f9}. Moreover, for uncertain fractional-order systems, \cite{f10,f11} proposed two adaptive backstepping control methods by utilizing FLSs. Employing the indirect Lyapunov method \cite{f12},  adaptive  fractional-order controllers were proposed for state and output feedback nonlinear systems  in \cite{f5,f6}, respectively.
It is notable that so far, very few results on the control of fractional-order interconnected systems have been obtained. In our previous work \cite{f17}, we developed a decentralized adaptive control method for a class of uncertain fractional-order interconnected  systems.

However, to the best of our knowledge, there are still no results on the tracking control problem for fractional-order systems with unknown control directions, especially for interconnected systems. The main difficulty is that the Nussbaum function technique, which is commonly used for integer-order systems with unknown control directions, cannot be directly used for fractional-order systems. Fractional-order dynamics present technical challenges to the solvability of the problem. First, for integer-order systems, the integer-order integral transformation is combined with the Nussbaum function to establish exponential stability. However, for fractional-order systems, one needs to utilize the fractional-order integral transformation and Nussbaum function to show Mittag--Leffler stability. The existence of fractional-order operator, Nussbaum and Mittag--Leffler functions will cause great difficulties in stability analysis. Second, fractional-order interconnected systems contain multiple unknown control directions. This will further complicate the problem since  dealing with multiple unknown control directions is not easy even with integer-order systems and only a few related results have been reported.

Motivated by the above considerations, the tracking control problem for fractional-order interconnected systems with unknown control directions will be investigated in this paper. A new decentralized adaptive fuzzy control method is proposed by using backstepping technique. The main contributions of the paper are  as follows.

\begin{enumerate}
  \item A novel fractional-order Nussbaum function technique is proposed to lay the foundation for the stability analysis of fractional-order systems with unknown control directions. This technique is much more general than those of existing works \cite{u4,u8} since it not only handles single/multiple unknown control directions but is also suitable for fractional/integer-order single/interconnected systems.

  \item As far as we know, this is the first work on decentralized control for fractional-order interconnected systems with unknown control directions.  In particular, unknown interactions among subsystems are bounded by nonlinear growth conditions. Smooth functions are introduced to compensate for  interconnections.

  \item FLSs are adopted to approximate the unknown nonlinear functions and fractional-order derivative of the virtual control law. This technique can handle  unknown uncertainties and reduce the computational burden of the controller considerably.
\end{enumerate}

The organization of the rest of this paper is as follows. Section II provides the preliminaries and problem formulation Section III proposes the new fractional-order Nussbaum function technique. The decentralized adaptive controller and its stability analysis are presented in Sections IV and V, respectively. Finally, Section VI provides the simulation results. Section VII concludes the paper.

\section{Preliminaries and problem formulation}
\subsection{Preliminaries}
Before designing the controller, we present some useful definitions and lemmas.
For a smooth function $f(t)$, define the following Caputo fractional-order derivative with order $\alpha$
\begin{equation}
{_0{\rm{{\cal D}}}^\alpha_t }f(t)=\frac{1}{\Gamma(n-\alpha)}\int_{0}^t\frac{f^{(n)}(\tau)}{(t-\tau)^{a-n+1}}d\tau,
\end{equation}
where ${_{0}{\rm{{\cal D}}}^\alpha_t }$ is the fractional-order differential operator and $\alpha\in[m-1,m)$, $m\in \mathbb N$, $\Gamma(\alpha)=\int_0^\infty x^{\alpha-1}e^{-x}dx$ denotes the Gamma function. To simplify the expression, ${_0{\rm{{\cal D}}}^\alpha_t }$ is denoted as ${\rm{{\cal D}}}^\alpha$. We obtain the following properties for Caputo's derivative.

\textbf{\emph{Lemma 1 {\cite{f0}}:}} For smooth functions $f(t)$, $h(t)$: $[t_0,\infty)\rightarrow\mathbb{R}$,
\begin{align*}
{\rm{{\cal L}}}\big({{\rm{{\cal D}}}^\alpha_t }(f(t)\big)&=s^\alpha F(s)-\sum\limits_{k=0} ^{n-1} s^{\alpha-k-1}f^{k}(0),\\
{{\rm{{\cal D}}}^\alpha_t }\big(af(t)+bh(t)\big)&=a{{\rm{{\cal D}}}^\alpha_t }f(t)+b{{\rm{{\cal D}}}^\alpha_t }h(t),\\
{{\rm{{\cal D}}}^\alpha_t }a&=0
\end{align*}
hold, where ${\rm{{\cal L}}}(\cdot)$ is the Laplace operator, $F(s)$ is the Laplace transform of $f(t)$, $a$ and $b$ are constants.

\textbf{\emph{Lemma 2 {\cite{f9}}:}} For a smooth function $x(t)\in\mathbb{R}$, one obtains
\begin{equation}
\frac{1}{2}{\rm{{\cal D}}}^\alpha\big(x^T(t)x(t)\big)\leq x^T(t){\rm{{\cal D}}}^\alpha x(t),\ t\geq t_0.
\end{equation}

Next, we will introduce the definition of the Mittag--Leffler function, which plays a key role in the stability analysis.

\textbf{\emph{Definition 1 {\cite{f0}}:}} Define the Mittag--Leffler function
\begin{equation}
E_{a,b}(\nu)=\sum\limits_{k=1}^\infty\frac{\nu^k}{\Gamma(ka+b)},\label{1}
\end{equation}
where   $\nu$ is a complex number and $a$ and $b$ are  two positive parameters. It is noteworthy that $E_{1,1}(\nu)=e^\nu$. The Laplace transform of \eqref{1} is
\begin{equation}
{\rm{{\cal L}}}\big\{t^{b-1}E_{a,b}(-\gamma t^a)\big\}=\frac{s^{a-b}}{s^a+\gamma},\gamma\in \mathbb R.\label{2}
\end{equation}

The Mittag--Leffler function has the following property.

\textbf{\emph{Lemma 3 {\cite{f0}}:}} Given real numbers $a\in(0,2)$, $b\in \mathbb R$, $\mu\in(\frac{a\pi}{2},\min\{\pi,a\pi\})$, we obtain
\begin{equation}
|E_{a,b}(\nu)|\leq\frac{\sigma}{1+|\nu|},
\end{equation}
where $\sigma>0$, $\mu\leq|\arg(e)|\leq\pi$, and $|\nu|\geq0$.\\

\subsection{Problem formulation}
Consider the following fractional-order interconnected nonlinear systems with $N$ subsystems:
\begin{equation}
\begin{aligned}
{{\rm{{\cal D}}}^\alpha }{x_{i,j}} &= g_{i,j}(t){x_{i,j+1}} + {\phi _{i,j}}({\bar x_{i,j}}) + {f_{i,j}( y)},\\
%
%
{{\rm{{\cal D}}}^\alpha }{x_{i,n_i}} &= g_{i,n_i}(t){u_i} + {\phi _{i,n_i}}({x_i}) + {d_i}(t) + {f_{i,n_i}( y)},\\
{y_i}& = {x_{i,1}},j=1,\ldots,n_i-1,\label{3}
\end{aligned}
\end{equation}
where $i=1,2,\ldots,N$. $\alpha \in (0,1)$ is the fractional order of the subsystem. ${x_i} = {\left[ {{x_{i,1}},{x_{i,2}}, \ldots ,{x_{i,n_i}}} \right]^T}\in \mathbb{R} ^ {n_i}$ and ${y_i}\in \mathbb{R}$ denote the states and output of the $i$th subsystem, respectively. ${\bar x_{i,j}} = {\left[{{x_{i,1}},{x_{i,2}}, \ldots ,{x_{i,j}}} \right] ^ T} $, $j=1,2,\ldots,n_i-1$, $ y=[y_1,\ldots,y_N]^T$. ${u_i}\in \mathbb{R}$ represents the control input. $d_i(t) \in \mathbb{R}$ is an unknown external disturbance. ${\phi _{i,j}}(\cdot) \in \mathbb{R}$, $j=1,2,\ldots,n_i$ represents unknown smooth nonlinear functions. $f_{i,j}(\cdot)$, $j=1,2,\ldots,n_i$ represents the unknown nonlinear interactions from other subsystems. $g_{i,j}(t)\neq0$, $j=1,2,\ldots,n_i$ represents the unknown time-varying control coefficient with an uncertain control direction.

The following assumptions are given for the above systems.

\textbf{\emph{Assumption 1:}} The unknown time-varying control coefficient $g_{i,j}$ is bounded in the closed intervals $I_{i,j}:=[\bar g_{i,j}^-,\bar g_{i,j}^+]$ with $0\notin I_{i,j}$, and the signs of $g_{i,j}$, $i\in[1,N]$, $j\in[1,n_i]$ are identical and unknown.

\textbf{\emph{Assumption 2:}} For $j=1,\ldots,n_i$, the nonlinear interactions $f_{i,j}(t, y)$ satisfy
\begin{equation*}
\big| f_{i,j}(t, y)\big|\leq \sum\limits_{q=1}^N  \beta_{i,j,q}\big|\psi_{i,j,q}(y_{q})\big|,
\end{equation*}
where $ \beta_{i,j,q}$ is an unknown positive constant and $\psi_{i,j,q}(y_q)$ is a known smooth function.

\textbf{\emph{Assumption 3:}} There is an unknown positive constant $\bar d_i$. For all $t\geq0$, it is established that $|d_i|\leq\bar d_i$.

\textbf{\emph{Assumption 4:}} The reference signal $y_{di}$  and its fractional-order derivative ${\rm{{\cal D}}}^\alpha y_{ri}$ are smooth, known, and bounded.

\textbf{\emph{Remark 1:}} The above assumptions are standard in controller design. Assumption 1 means that the time-varying control coefficients $g_{i,j}$ are either strictly positive or strictly negative with the same unknown sign. This assumption is adapted from \cite{u8}. Assumptions 3 and 4 are similar to those in \cite{f10,f11}. Assumption 2 implies that the unknown interactions $f_{i,j}$ are bounded by a relaxed nonlinear growth condition. This bounding condition is reasonable and can be found in \cite{interaction}.

\textbf{\emph{Remark 2:}} A fractional-order model can characterize the dynamics of many practical applications in engineering well, such as servo motor systems, robotic manipulators, and power systems. Many valuable results on fractional-order modeling and control have been reported (see refs. \cite{fa1,fa4,motor_f,fa6} for details). For example, \cite{motor_f} established a fractional-order nonlinear model of a permanent magnet synchronous motor. Simulations have verified the advantages of the proposed fractional-order control scheme. Meanwhile, when some single fractional-order systems are linked or coupled together, a fractional-order interconnected system is constructed. For instance, several motors are linked by gear, spring or transmission networks, etc. In addition, some fractional-order multiple-input multiple-output (MIMO) systems \cite{motor_f,motor_MIMO} can be expressed as fractional-order interconnected systems (see Example 2 in Section V for more details).

Our control aim  is to develop a decentralized adaptive controller such that all the signals are bounded and the tracking errors $y_i-y_{ri}$, $i=1,2,\ldots,N$ can converge to a small neighborhood of origin.

\subsection{Fuzzy logic systems}
The following FLSs \cite{Fuzzy} are utilized to approximate the unknown nonlinear functions in this paper. A FLS contains four ingredients: the knowledge base, the fuzzifier,  the fuzzy inference engine, and the defuzzifier. The knowledge base is constructed with the following fuzzy linguistic rules:
\begin{align*}
R^k:\ &{If}\ x_1\ is\ F^k_1\ and\ x_2\ {is}\ F_2^k\ and\ \ldots\ and\ x_n\ \emph{is}\ F_n^k,\\
&{then}\ y\ {is}\ P^k,\ k=1,2,\ldots,N,
\end{align*}
where $x=(x_1,\ldots,x_n)^T$ and $y$ denote the input and output of the FLSs, respectively and $F_i^k$ and $P_i^k$ are fuzzy sets in $\mathbb{R}$, $i=1,2,\ldots,n$.

Utilizing a singleton fuzzifier, product inference and center average defuzzification, FLSs can be expressed as
\begin{equation}
y(x)=\frac{\sum_{k=1}^N\bar y_k\prod_{i=1}^n\mu_{F_i^k}(x_i)}{\sum^N_{k=1}\big(\prod_{i=1}^n\mu_{F_i^k}(x_i)\big)},\label{71}
\end{equation}
where $\mu_{F_i^k}(x_i)$ is the membership function, $\bar y_k= \max_{y\in R} \mu_{P^k}(y)$. Here, \eqref{71} can be transformed into
\begin{equation}
{y(x)}=\theta^T\varphi(x),\label{4}
\end{equation}
where $\theta=[\theta_1,\theta_2,\ldots,\theta_N]^T=[\bar y_1,\bar y_2,\ldots,\bar y_N]^T$, $\varphi(x)=\big[\varphi_1(x),\varphi_2(x),\ldots,\varphi_N(x)\big]^T$ with
\begin{equation}
\varphi_k(x)=\frac{\prod_{i=1}^n\mu_{F_i^k}(x_i)}{\sum^N_{k=1}\big(\prod_{i=1}^n\mu_{F_i^k}(x_i)\big)},k=1,2,\ldots,N.
\end{equation}

For FLSs, we obtain an important property.

\textbf{\emph{Lemma 4 {\cite{Fuzzy}}:}} Assuming that $f(x):\Omega\rightarrow\mathbb{R}$ is a continuous function with $\Omega$ being a compact set, an FLS exists such that
\begin{equation*}
\mathop{\sup}\limits_{x\in\Omega}\big|f (x)-\theta^T\varphi(x)\big|\leq \epsilon
\end{equation*}
with an arbitrarily small constant $\epsilon>0$.

\section{Novel fractional-order Nussbaum function technique}
In this section, we propose a novel fractional-order Nussbaum function technique to address the unknown direction of time-varying control coefficients $g_{i,j}$ in \eqref{3}.

The Nussbaum function $\mathcal{N}(\cdot)$ has different forms, which is intended to satisfy the properties shown as follows
\begin{gather}
\lim\limits_{x\rightarrow\infty}\sup\int\limits_{x_0}^x \mathcal{N}(\delta)d\delta=+\infty,\nonumber\\
\lim\limits_{x\rightarrow\infty}\inf\int\limits_{x_0}^x \mathcal{N}(\delta)d\delta=-\infty.
\end{gather}

Here, we choose the Nussbaum function as
\begin{equation}
\mathcal{N}(\delta)=e^{\delta^2}\sin{\frac{\delta}{2}\pi}.\label{5}
\end{equation}

Based on the Nussbaum function, we will present the key theorem for stability analysis.

\textbf{\emph{Theorem 1:}} Let $V(\cdot)$ and $\delta_i(\cdot)$, $i=1,\ldots,n$ be smooth functions defined on $[0,t_f)$ with $V(t)\geq 0$, $\forall t\in[0,t_f)$  and all $g_i(t)$, $i=1,\ldots,n$ be unknown time-varying parameters that have the same sign and satisfy $g_i\in I_i:=[\bar g_i^-,\bar g_i^+]$ with $0\notin I_i$. If the following inequality holds:
\begin{equation}
{{\rm{{\cal D}}}^\alpha }V(t)\leq -\lambda V+\sum\limits_{i=1}^n \big(g_i \mathcal{N}(\delta_i)\dot\delta_i+\dot\delta_i\big) + \zeta,\label{6}
\end{equation}
where $\lambda$ and $\zeta$ are positive constants, then $V(t)$, $\delta_i(t)$ must be bounded on $[0,t_f)$ for $i=1,2,\ldots,n$.

\textbf{\emph{Proof:}} See Appendix A.

\textbf{\emph{Remark 3:}} The form of inequality \eqref{6} is the $\alpha$th-order derivative of the Lyapunov function. Through Laplace transformation,  $V(t)$ is described  as
\begin{align}
V(t)\leq &\sum\limits_{i = 1}^n\int\limits_0^tg_i\mathcal{N}(\delta_i)\dot\delta_i E_{\alpha,\alpha}(-\lambda(t-\varsigma)^\alpha)(t-\varsigma)^{\alpha-1}d\varsigma\nonumber\\
& +\sum\limits_{i = 1}^n\int\limits_0^t\dot\delta_i E_{\alpha,\alpha}(-\lambda(t-\varsigma)^\alpha)(t-\varsigma)^{\alpha-1}d\varsigma \nonumber\\
& + H.\label{72}
\end{align}

According to Theorem 1, it can be found that $\int\limits_0^tg_i\mathcal{N}(\delta_i)\dot\delta_i E_{\alpha,\alpha}(-\lambda(t-\varsigma)^\alpha)(t-\varsigma)^{\alpha-1}d\varsigma$ and $\int\limits_0^t\dot\delta_i E_{\alpha,\alpha}(-\lambda(t-\varsigma)^\alpha)(t-\varsigma)^{\alpha-1}d\varsigma$ are bounded. Moreover, $t_f$ is $\infty$ because the solution is bounded.

\textbf{\emph{Remark 4:}}
The Nussbaum function technique is widely used for the control of integer-order systems with unknown control directions. The stability analysis is mainly based on the integer-order integral transformation and exponential stability. However, the method is not suitable for fractional-order systems because the stability analysis of fractional-order systems relies on  fractional-order integral transformation and the Mittag--Leffler function (see \eqref{72}). The Mittag--Leffler function is an indispensable tool for analyzing the stability of fractional-order systems, which makes the analysis more complicated. It is challenging to show the boundedness of all the signals based on the Nussbaum function. Therefore, we present Theorem 1 which successfully overcomes the above difficulties and lays the foundation for the controller design in Section IV.

\textbf{\emph{Remark 5:}} Theorem 1 is the stability criterion for multiple Nussbaum functions, which can be adopted for fractional-order interconnected systems. Note that it can also be  adopted for a single Nussbaum function ($n=1$) or integer-order systems ($\alpha=1$). In fact, Theorem 1 is new even when $n=1$.

\section{Decentralized adaptive controller design}
The change in coordinates is defined as
\begin{align}
& z_{i,1}=x_{i,1}-y_{ri},\label{75}\\
& z_{i,j}=x_{i,j}-\tau_{i,j-1},j=2,3,\ldots,n_i,\label{7}
\end{align}
where $\tau_{i,j}$ is the virtual control law. Then, we follow the following controller design procedures.

\textbf{\emph{Step} \emph{1}\emph{:}} From equation \eqref{75}, the $\alpha$th-order derivative of $z_{i,1}$ is expressed as
\begin{equation}
\begin{aligned}
{{\rm{{\cal D}}}^\alpha }{z_{i,1}}{}&={{\rm{{\cal D}}}^\alpha }{x_{i,1}}-{{\rm{{\cal D}}}^\alpha }{y_{ri}}\\
&=g_{i,1}{x_{i,2}} + {\phi _{i,1}} + {f_{i,1}}-{{\rm{{\cal D}}}^\alpha }{y_{ri}}.\label{8}
\end{aligned}
\end{equation}

According to Lemma 4, we use the FLSs to approximate the unknown smooth function ${\phi _{i,1}}({\bar x_{i,1}})$:
\begin{equation}
 {\hat{\phi} _{i,1}}({\bar x_{i,1}},\theta_{i,1}^T)=\theta_{i,1}^T\varphi_{i,1}(\bar x_{i,1}).\label{9}
\end{equation}

In the bounded compact sets $\Omega_{i,1}$ and $U_{i,1}$, the ideal parameter vector $\theta_{i,1}^\ast$ is defined by
\begin{equation*}
\theta_{i,1}^\ast=\arg\mathop{\min }\limits_{{\theta _{i,1}\in\Omega_{i,1}}}\Big[\mathop{\sup}\limits_{{\bar x_{i,1}\in U_{i,1}}}\big|{\phi _{i,1}}-{\hat{\phi} _{i,1}}({\bar x_{i,1}},\theta_{i,1})\big|\Big].
\end{equation*}

The optimal approximation errors are given as
\begin{equation}
\varepsilon_{i,1}({\bar x_{i,1}})={\phi _{i,1}}-{\hat{\phi} _{i,1}}.\label{10}
\end{equation}

Note that an unknown constant $\bar \varepsilon_{i,1}>0$ exists such that $\big|\varepsilon_{i,1}({\bar x_{i,1}})\big|\leq \bar \varepsilon_{i,1}$.
Then, from \eqref{8}--\eqref{10}, one immediately obtains
\begin{equation}
\begin{aligned}
{{\rm{{\cal D}}}^\alpha }{z_{i,1}}{}&=g_{i,1}z_{i,2} + g_{i,1}\tau_{i,1} + {\phi _{i,1}} - {\hat{\phi} _{i,1}} \\
&\ \ \ \ + {\hat{\phi} _{i,1}}+ {f_{i,1}}-{{\rm{{\cal D}}}^\alpha }{y_{ri}}\\
&=g_{i,1}z_{i,2} + g_{i,1}\tau_{i,1} + \varepsilon_{i,1}+{{\tilde \theta }_{i,1}}^T\varphi_{i,1} \\
&\ \ \ \ + \theta_{i,1}^T\varphi_{i,1}+f_{i,1}-{{\rm{{\cal D}}}^\alpha }{y_{ri}}.\label{11}
\end{aligned}
\end{equation}

The virtual control law $\tau_{i,1}$ is designed as
\begin{equation}
\begin{aligned}
\tau_{i,1}&=\mathcal{N}(\delta_{i,1})\eta_{i,1},\\
\eta_{i,1}&=c_{i,1}z_{i,1}  +k_{i,1}z_{i,1} + l_{i,1}z_{i,1}+ \theta_{i,1}^T\varphi_{i,1}\\
&\ \ \ \  + \hat \mu_i h_i - {{\rm{{\cal D}}}^{\alpha} }{y_{ri}},\\
\dot \delta_{i,1}&=z_{i,1}\eta_{i,1},\\
h_i&=\frac{2z_{i,1}}{z_{i,1}^2+\varpi_i}\sum\limits_{q=1}^N \sum\limits_{j=1}^n \psi_{q,j,i}^2(y_i),\label{12}
\end{aligned}
\end{equation}
where $c_{i,1}>\frac{1}{4}$, $k_{i,1}$, $l_{i,1}$ and $\varpi_i$ are positive design parameters; $h_i$ is a constructed auxiliary smooth function for compensating for the interactions; and $\hat \mu_i$ denotes the estimation of  $\mu_i$ which is the upper bound of $\sum\limits_{q = 1}^n\sum\limits_{j=1}^N\frac{1}{{4{l_{q,j}}}}\beta_{q,j,i}^2$. Substituting \eqref{12} into \eqref{11} and introducing a variable $\eta_{i,1}$ yields
\begin{equation*}
\begin{aligned}
{{\rm{{\cal D}}}^\alpha }{z_{i,1}}
&=g_{i,1}z_{i,2} + g_{i,1}\mathcal{N}(\delta_{i,1})\eta_{i,1} +\eta_{i,1}-\eta_{i,1}+ \varepsilon_{i,1} \\
&\ \ \ \ +{{\tilde \theta }_{i,1}}^T\varphi_{i,1}+ \theta_{i,1}^T\varphi_{i,1}+f_{i,1}-{{\rm{{\cal D}}}^\alpha }{y_{ri}}\\
&=g_{i,1}z_{i,2}+g_{i,1}\mathcal{N}(\delta_{i,1})\eta_{i,1}+\eta_{i,1}-c_{i,1}z_{i,1}   \\
& \ \ \ \ -k_{i,1}z_{i,1}+\varepsilon_{i,1}- l_{i,1}z_{i,1}+f_{i,1}\\
&\ \ \ \ - \hat \mu_i h_i +{{\tilde \theta }_{i,1}}^T\varphi_{i,1}.
\end{aligned}
\end{equation*}

By multiplying both sides by $z_{i,1}$ and using Young's inequality $ ab\leq a^2 + \frac{1}{4}b^2$, one obtains
\begin{align}
z_{i,1}{{\rm{{\cal D}}}^\alpha }{z_{i,1}}
&=g_{i,1}z_{i,1}z_{i,2} + g_{i,1}\mathcal{N}(\delta_{i,1})\dot\delta_{i,1} +\dot\delta_{i,1}-c_{i,1}z_{i,1}^2\nonumber\\
&\ \ \ \  -k_{i,1}z_{i,1}^2 +z_{i,1}\varepsilon_{i,1} -l_{i,1}z_{i,1}^2+ z_{i,1}f_{i,1}  \nonumber\\
&\ \ \ \ -\hat \mu_i z_{i,1}h_i+{{\tilde \theta }_{i,1}}^T\varphi_{i,1}z_{i,1}\nonumber\\
&\leq g_{i,1}\mathcal{N}(\delta_{i,1})\dot\delta_{i,1} +\dot\delta_{i,1}-\bar c_{i,1}z_{i,1}^2+g_{i,1}^2z_{i,2}^2\nonumber\\
&\ \ \ \ + \frac{1}{4k_{i,1}}\varepsilon_{i,1}^2+\frac{1}{4l_{i,1}}f_{i,1}^2-\hat \mu_i z_{i,1}h_i\nonumber\\
&\ \ \ \ +{{\tilde \theta }_{i,1}}^T\varphi_{i,1}z_{i,1},\label{13}
\end{align}
where $\bar c_{i,1}=c_{i,1}-\frac{1}{4}$.

Construct the Lyapunov function as
\begin{equation}
V_{i,1}= \frac {1}{2}z_{i,1}^2 + \frac{1}{2}{\tilde \theta }_{i,1}^T\Lambda_{i,1}^{-1}{\tilde \theta }_{i,1}+\frac{1}{2\gamma_{i,1}}\tilde\mu_{i}^2,\label{15}
\end{equation}
where $\Lambda_{i,1}$ is a positive-definite matrix and $\gamma_{i,1}$ is a positive design parameter. From Lemmas 1--2 and \eqref{15}, differentiating $V_{i,1}$ with the $\alpha$th order yields
\begin{equation}
\begin{aligned}
{{\rm{{\cal D}}}^\alpha }V_{i,1}&=\frac{1}{2}{{\rm{{\cal D}}}^\alpha }z_{i,1}^2 + \frac{1}{2}{{\rm{{\cal D}}}^\alpha }{\tilde \theta }_{i,1}^T\Lambda_{i,1}^{-1}{\tilde \theta }_{i,1}+\frac{1}{2\gamma_{i,1}}{{\rm{{\cal D}}}^\alpha }\tilde\mu_{i}^2\\
&\leq z_{i,1}{{\rm{{\cal D}}}^\alpha }z_{i,1} - {\tilde \theta }_{i,1}^T\Lambda_{i,1}^{-1}{{\rm{{\cal D}}}^\alpha }\theta_{i,1}-\frac{1}{\gamma_{i,1}}\tilde\mu_{i}{{\rm{{\cal D}}}^\alpha }\hat\mu_{i}.\label{16}
\end{aligned}
\end{equation}

Substituting \eqref{13} into \eqref{16} yields
\begin{equation}
\begin{aligned}
{{\rm{{\cal D}}}^\alpha }V_{i,1}
\leq &-\bar c_{i,1}z_{i,1}^2+g_{i,1}\mathcal{N}(\delta_{i,1})\dot\delta_{i,1} +\dot\delta_{i,1}+g_{i,1}^2z_{i,2}^2\\
& + \frac{1}{4k_{i,1}}\bar\varepsilon_{i,1}^2 +\frac{1}{4l_{i,1}}f_{i,1}^2- \mu_i z_{i,1}h_i \\
& -{\tilde \theta }_{i,1}^T\Lambda_{i,1}^{-1}({{\rm{{\cal D}}}^\alpha }\theta_{i,1}-\Lambda_{i,1}\varphi_{i,1}z_{i,1})\\
& -\frac{1}{\gamma_{i,1}}\tilde\mu_{i}({{\rm{{\cal D}}}^\alpha }\hat\mu_{i}-\gamma_{i,1}z_{i,1}h_i).\label{17}
\end{aligned}
\end{equation}

Design the update laws ${{\rm{{\cal D}}}^\alpha }\theta_{i,1}$ and ${{\rm{{\cal D}}}^\alpha }\hat\mu_{i}$ as
\begin{align}
{{\rm{{\cal D}}}^\alpha }\theta_{i,1} &= \Lambda_{i,1}\varphi_{i,1}z_{i,1} - \rho_{i,1}\theta_{i,1},\label{18}\\
{{\rm{{\cal D}}}^\alpha }\hat\mu_{i} &= \gamma_{i,1}z_{i,1}h_i-\gamma_{i,2}\hat\mu_i.\label{19}
\end{align}

Substituting \eqref{18} and \eqref{19} into \eqref{17} yields
\begin{equation*}
\begin{aligned}
{{\rm{{\cal D}}}^\alpha }V_{i,1}
\leq& -\bar c_{i,1}z^2_{i,1}  +g_{i,1}\mathcal{N}(\delta_{i,1})\dot\delta_{i,1} +\dot\delta_{i,1}+g_{i,1}^2z_{i,2}^2\\
& + \frac{1}{4k_{i,1}}\bar\varepsilon_{i,1}^2 +\frac{1}{4l_{i,1}}f_{i,1}^2- \mu_i z_{i,1}h_i  \\
 & +\rho_{i,1}{\tilde \theta }_{i,1}^T\Lambda_{i,1}^{-1}\theta_{i,1}^\ast- \rho_{i,1}{\tilde \theta }_{i,1}^T\Lambda_{i,1}^{-1}{\tilde \theta }_{i,1}\\
&  +\frac{\gamma_{i,2}}{\gamma_{i,1}}\tilde\mu_{i}\mu_i-\frac{\gamma_{i,2}}{\gamma_{i,1}}\tilde\mu_{i}^2.\\
\end{aligned}
\end{equation*}

Through Young's inequality, we obtain
\begin{equation}
\begin{aligned}
{{\rm{{\cal D}}}^\alpha }V_{i,1}
\leq &-\bar c_{i,1}z^2_{i,1}  +g_{i,1}\mathcal{N}(\delta_{i,1})\dot\delta_{i,1} +\dot\delta_{i,1}+g_{i,1}^2z_{i,2}^2\\
& + \frac{1}{4k_{i,1}}\bar\varepsilon_{i,1}^2 +\frac{1}{4l_{i,1}}f_{i,1}^2- \mu_i z_{i,1}h_i  \\
& +\frac{1}{2}\rho_{i,1}{ \theta }_{i,1}^{\ast T}\Lambda_{i,1}^{-1}\theta_{i,1}^\ast- \frac{1}{2}\rho_{i,1}{\tilde \theta }_{i,1}^T\Lambda_{i,1}^{-1}{\tilde \theta }_{i,1}\\
& +\frac{\gamma_{i,2}}{2\gamma_{i,1}}\mu_{i}^2-\frac{\gamma_{i,2}}{2\gamma_{i,1}}\tilde\mu_{i}^2\\
\leq &-\lambda_{i,1}V_{i,1}+g_{i,1}^2z_{i,2}^2+\zeta_{i,1},\label{20}
\end{aligned}
\end{equation}
where $\lambda_{i,1}=\min \{2\bar c_{i,1},\rho_{i,1},\gamma_{i,2}\}$ and $\zeta_{i,1}=g_{i,1}\mathcal{N}(\delta_{i,1})\dot\delta_{i,1} +\dot\delta_{i,1}+\frac{1}{4k_{i,1}}\bar\varepsilon_{i,1}^2+\frac{1}{{4{l_{i,1}}}}f_{i,1}^2-\mu_i z_{i,1}h_i+\frac{\rho_{i,1}}{2}\theta_{i,1}^{\ast T}\Lambda_{i,1}^{-1}\theta_{i,1}^\ast+\frac{\gamma_{i,2}}{2\gamma_{i,1}}\mu_{i}^2$.

\textbf{\emph{Step} \emph{j} }( $\emph{j}={{2}},{\ldots},{n}_{{i-1}}$)\emph{\textbf{:}} From \eqref{7}, we obtain
\begin{equation*}
\begin{aligned}
{{\rm{{\cal D}}}^\alpha }{z_{i,j}}{}&=g_{i,j}{x_{i,j+1}} + {\phi _{i,j}} + {f_{i,j}}-{{\rm{{\cal D}}}^\alpha }\tau_{i,j-1} \\
&=g_{i,j}{z_{i,j+1}} + g_{i,j}\tau_{i,j} + {\phi _{i,j}}+ {f_{i,j}} -{{\rm{{\cal D}}}^{\alpha} }\tau_{i,j-1}\\
&\ \ \ \  +g_{i,j-1}^2z_{i,j}-g_{i,j-1}^2z_{i,j}\\
&=g_{i,j}{z_{i,j+1}} + g_{i,j}\tau_{i,j} + {\Phi _{i,j}}+ {f_{i,j}}-g_{i,j-1}^2z_{i,j},
\end{aligned}
\end{equation*}
where $g_{i,j-1}^2z_{i,j}$ is an auxiliary term and ${\Phi _{i,j}}={\phi _{i,j}}-{{\rm{{\cal D}}}^\alpha }\tau_{i,j-1}+g_{i,j-1}^2z_{i,j}$ is the unknown nonlinear function. Using FLSs to approximate ${\Phi _{i,j}}$ yields
\begin{equation}
\begin{aligned}
{{\rm{{\cal D}}}^\alpha }{z_{i,j}}{}& =g_{i,j}{z_{i,j+1}} + g_{i,j}\tau_{i,j} +\Phi_{i,j} - {\hat{\Phi} _{i,j}}\\
&\ \ \ \  +{\hat{\Phi} _{i,j}}+ {f_{i,j}}-g_{i,j-1}^2z_{i,j}\\
&=g_{i,j}{z_{i,j+1}} + g_{i,j}\tau_{i,j} +\varepsilon_{i,j}+{{\tilde \theta}_{i,j}^T}\varphi_{i,j} \\
&\ \ \ \  + \theta_{i,j}^T\varphi_{i,j}+f_{i,j}-g_{i,j-1}^2z_{i,j},\label{21}
\end{aligned}
\end{equation}
where the approximation error is $|\varepsilon_{i,j}|\leq\bar\varepsilon_{i,j}$.

The virtual controller $\tau_{i,j}$ and adaptation law ${{\rm{{\cal D}}}^\alpha }\theta_{i,j}$ are designed as
\begin{align}
\tau_{i,j}&=\mathcal{N}(\delta_{i,j})\eta_{i,j},\label{22}\\
\eta_{i,j}&=c_{i,j}z_{i,j}+k_{i,j}z_{i,j}+l_{i,j}z_{i,j}+\theta_{i,j}^T\varphi_{i,j},\nonumber\\ \nonumber
\dot \delta_{i,j}&=z_{i,j}\eta_{i,j},\nonumber\\
{{\rm{{\cal D}}}^\alpha }\theta_{i,j}& = \Lambda_{i,j}\varphi_{i,j}z_{i,j} - \rho_{i,j}\theta_{i,j},\label{23}
\end{align}
where $c_{i,j}>\frac{1}{4}$, $k_{i,j}$ and $l_{i,j}$ are positive constants and $\Lambda_{i,j}$ is a positive-definite matrix. By using \eqref{21}--\eqref{22} and multiplying both sides by $z_{i,j}$, one obtains
\begin{align}
z_{i,j}{{\rm{{\cal D}}}^\alpha }{z_{i,j}}&\leq g_{i,j}\mathcal{N}(\delta_{i,j})\dot\delta_{i,j}+\dot\delta_{i,j} -\bar c_{i,j}z_{i,j}^2+g_{i,j}^2z_{i,j+1}^2\nonumber\\
&\ \ \ \  +\frac{1}{4k_{i,j}}\varepsilon_{i,j}^2+\frac{1}{4l_{i,j}}f_{i,j}^2+{{\tilde \theta }_{i,j}}^T\varphi_{i,j}z_{i,j}\nonumber\\
&\ \ \ \  -g_{i,j-1}^2z_{i,j}^2,\label{24}
\end{align}
where $\bar c_{i,j}= c_{i,j}-\frac{1}{4}$.
Select the Lyapunov function as
\begin{equation}
V_{i,j}=V_{i,j-1}+ \frac {1}{2}z_{i,j}^2 + \frac{1}{2}{\tilde \theta }_{i,j}^T\Lambda_{i,j}^{-1}{\tilde \theta }_{i,j}.\label{25}
\end{equation}

Then, from \eqref{20} and \eqref{23}--\eqref{25}, we obtain
\begin{align}
{{\rm{{\cal D}}}^\alpha }V_{i,j}\leq & -\lambda_{i,j-1}V_{i,j-1}+\zeta_{i,j-1}-\bar c_{i,j}z_{i,j}^2\nonumber+g_{i,j}^2z_{i,j+1}^2\\
&  +g_{i,j}\mathcal{N}(\delta_{i,j})\dot\delta_{i,j}+\dot\delta_{i,j} +\frac{1}{4k_{i,j}}\bar\varepsilon_{i,j}^2\nonumber+\frac{1}{4l_{i,j}}f_{i,j}^2\\
&   + {\tilde \theta }_{i,j}^T\Lambda_{i,j}^{-1}({{\rm{{\cal D}}}^\alpha }{\theta }_{i,j}-\Lambda_{i,j}^{-1}\varphi_{i,j}z_{i,j})\nonumber\\
\leq & -\lambda_{i,j-1}V_{i,j-1}+\zeta_{i,j-1}-\bar c_{i,j}z_{i,j}^2\nonumber+g_{i,j}^2z_{i,j+1}^2\\
&  +g_{i,j}\mathcal{N}(\delta_{i,j})\dot\delta_{i,j}+\dot\delta_{i,j} +\frac{1}{4k_{i,j}}\bar\varepsilon_{i,j}^2\nonumber +\frac{1}{4l_{i,j}}f_{i,j}^2\\
&   + \frac{\rho_{i,j}}{2}{ \theta }_{i,j}^{\ast T}\Lambda_{i,j}^{-1}{\theta }_{i,j}^\ast-\frac{\rho_{i,j}}{2}{ \tilde\theta }_{i,j}^{T}\Lambda_{i,j}^{-1}{\tilde \theta }_{i,j}\nonumber\\
\leq & -\lambda_{i,j}V_{i,j}+g_{i,j}^2z_{i,j+1}^2+\zeta_{i,j},\label{26}
\end{align}
where $\lambda_{i,j}=\min\{\lambda_{i,j-1},2\bar c_{i,j},\rho_{i,j}\}$ and $\zeta_{i,j}=\zeta_{i,j-1}++g_{i,j}N(\delta_{i,j})\dot\delta_{i,j}+\dot\delta_{i,j}+\frac{1}{4k_{i,j}}\bar\varepsilon_{i,j}^2+\frac{1}{{4{l_{i,j}}}}f_{i,j}^2+\frac{\rho_{i,j}}{2}\theta_{i,j}^{\ast T}\Lambda_{i,j}^{-1}\theta_{i,j}^\ast$.

\textbf{\emph{Step} $\emph{n}_{\emph{i}}$\emph{:}} Similar to the above steps, the $\alpha$th-order derivative of $z_{i,n_i}$ is
\begin{equation}
\begin{aligned}
{{\rm{{\cal D}}}^\alpha }{z_{i,n_i}}{}&=g_{i,n_i}u_i + {\phi _{i,n_i}} +d_i +{f_{i,n_i}} -{{\rm{{\cal D}}}^{\alpha} }\tau_{i,n_i-1}\\
&=g_{i,n_i}u_i + {\Phi _{i,n_i}} +d_i +{f_{i,n_i}} -g_{i,n_i-1}^2z_{i,n_i}\\
&=g_{i,n_i}u_i + \varepsilon_{i,n_i}+{{\tilde \theta}_{i,n_i}^T}\varphi_{i,n_i} +{{ \theta}_{i,n_i}^T}\varphi_{i,n_i}\\
&\ \ \ \ +d_i +{f_{i,n_i}} -g_{i,n_i-1}^2z_{i,n_i},\label{27}
\end{aligned}
\end{equation}
where the unknown nonlinear function ${\Phi _{i,n_i}}={\phi _{i,n_i}}-{{\rm{{\cal D}}}^\alpha }\tau_{i,n_i-1}+g_{i,n_i-1}^2z_{i,n_i}$ is approximated by FLSs, and the approximation error is $|\varepsilon_{i,n_i}|\leq\bar\varepsilon_{i,n_i}$.

Design the control input $u_i$ as
\begin{align}
u_i&= \mathcal{N}(\delta_{i,n_i})\eta_{i,n_i}, \label{28}\\
\eta_{i,n_i}&=c_{i,n_i}z_{i,n_i}+k_{i,n_i}z_{i,n_i}+l_{i,n_i}z_{i,n_i}\nonumber\\
&\ \ \ \ +{{ \theta}_{i,n_i}^T}\varphi_{i,n_i}+b_iz_{i,n_i},\nonumber\\
\dot\delta_{i,n_i}&=z_{i,n_i}\eta_{i,n_i}.\nonumber
\end{align}

Invoking \eqref{28}, multiplying both sides by $z_{i,n_i}$ and using Young's equality, \eqref{27} is denoted as
%
\begin{equation}
\begin{aligned}
z_{i,n_i}{{\rm{{\cal D}}}^\alpha }{z_{i,n_i}}{}
&\leq g_{i,n_i}\mathcal{N}(\delta_{i,n_i})\dot\delta_{i,n_i}+\dot\delta_{i,n_i}-c_{i,n_i}z_{i,n_i}^2\\
&\ \ \ \ +\frac{1}{4k_{i,n_i}}\varepsilon_{i,n_i}^2+\frac{1}{4l_{i,n_i}}f_{i,n_i}^2+\frac{1}{4b_i}d_i^2\\
&\ \ \ \ +{{\tilde \theta}_{i,n_i}^T}\varphi_{i,n_i}z_{i,n_i}-g_{i,n_i-1}^2z_{i,n_i}^2.\label{30}
\end{aligned}
\end{equation}

Consider
\begin{equation}
\begin{aligned}
V_{i,n_i}=&V_{i,n_i-1}+ \frac {1}{2}z_{i,n_i}^2 + \frac{1}{2}{\tilde \theta }_{i,n_i}^T\Lambda_{i,n_i}^{-1}{\tilde \theta }_{i,n_i},\label{31}
\end{aligned}
\end{equation}
where $\Lambda_{i,n_i}$ is a positive-definite matrix. Then, according to \eqref{30} and \eqref{31},
\begin{equation}
\begin{aligned}
{{\rm{{\cal D}}}^\alpha }V_{i,n_i}
&\leq -\lambda_{i,n_i-1}V_{i,n_i-1} +\zeta_{i,n_i-1}-c_{i,n_i}z_{i,n_i}^2\\
&\ \ \ \ +g_{i,n_i}\mathcal{N}(\delta_{i,n_i})\dot\delta_{i,n_i}+\dot\delta_{i,n_i}+\frac{1}{4k_{i,n_i}}\bar\varepsilon_{i,n_i}^2\\
&\ \ \ \ +\frac{1}{4l_{i,n_i}}f_{i,n_i}^2 +\frac{1}{4b_i}\bar {d}_i^2\\
&\ \ \ \ -{\tilde \theta }_{i,n_i}^T\Lambda_{i,n_i}^{-1}({{\rm{{\cal D}}}^\alpha }\theta_{i,n_i}-\Lambda_{i,n_i}\varphi_{i,n_i}z_{i,n_i}).\label{32}
\end{aligned}
\end{equation}

Design the update law ${{\rm{{\cal D}}}^\alpha }\theta_{i,n_i}$ as
\begin{gather}
{{\rm{{\cal D}}}^\alpha }\theta_{i,n_i}=\Lambda_{i,n_i}\varphi_{i,n_i}z_{i,n_i} - \rho_{i,n_i}\theta_{i,n_i}.\label{33}
\end{gather}

Substituting \eqref{33} into \eqref{32} yields
\begin{align}
{{\rm{{\cal D}}}^\alpha }V_{i,n_i}\nonumber
\leq & -\lambda_{i,n_i-1}V_{i,n_i-1} +\zeta_{i,n_i-1}-c_{i,n_i}z_{i,n_i}^2\\
&  \nonumber +g_{i,n_i}\mathcal{N}(\delta_{i,n_i})\dot\delta_{i,n_i}+\dot\delta_{i,n_i}+\frac{1}{4k_{i,n_i}}\bar\varepsilon_{i,n_i}^2\\
&  \nonumber+\frac{1}{4l_{i,n_i}}f_{i,n_i}^2+\frac{1}{4b_i}\bar {d}_i^2+\frac{\rho_{i,n_i}}{2}\theta_{i,n_i}^{\ast T}\Lambda_{i,n_i}^{-1}\theta_{i,n_i}^\ast\\
&  \nonumber- \frac{\rho_{i,n_i}}{2}{\tilde \theta }_{i,n_i}^T\Lambda_{i,n_i}^{-1}{\tilde \theta }_{i,n_i}\\
\leq & -\lambda_{i,n_i}V_{i,n_i}+\zeta_{i,n_i},\label{34}
\end{align}
where $\lambda_{i,n_i}=\min\{\lambda_{i,n_i-1},2c_{i,n_i},\rho_{i,n_i}\}$ and
\begin{gather*}
\begin{aligned}
\zeta_{i,n_i}&=\zeta_{i,n_i-1}+g_{i,n_i}\mathcal{N}(\delta_{i,n_i})\dot\delta_{i,n_i}+\dot\delta_{i,n_i}+\frac{1}{4k_{i,n_i}}\bar\varepsilon_{i,n_i}^2\\
&\ \ \ \ +\frac{1}{{4{l_{i,n_i}}}}f_{i,n_i}^2+\frac{1}{4b_i}\bar {d}_i^2+\frac{\rho_{i,n_i}}{2}\theta_{i,n_i}^{\ast T}\Lambda_{i,n_i}^{-1}\theta_{i,n_i}^\ast\\
&=\sum\limits_{j = 1}^{n_i} \big(g_{i,j}\mathcal{N}(\delta_{i,j})\dot\delta_{i,j}+\dot\delta_{i,n_i}\big)+\sum\limits_{j = 1}^{n_i}\frac{1}{4k_{i,j}}\bar\varepsilon_{i,j}^2\\
&\ \ \ \ +\sum\limits_{j = 1}^{n_i}\frac{1}{{4{l_{i,j}}}}f_{i,j}^2+\frac{1}{4b_i}\bar {d}_i^2+\sum\limits_{j = 1}^{n_i}\frac{\rho_{i,j}}{2}\theta_{i,j}^{\ast T}\Lambda_{i,j}^{-1}\theta_{i,j}^\ast\\
&\ \ \ \ -\mu_i z_{i,1}h_i+\frac{\gamma_{i,2}}{2\gamma_{i,1}}\mu_{i}^2.
\end{aligned}
\end{gather*}

\section{Stability analysis}
Using the controller designed above, we present the main results.

\textbf{\emph{Theorem 2:}}
Consider the fractional-order interconnected system  \eqref{3} with unknown control directions satisfying Assumptions 1--4, the control input \eqref{28}, and the parameter updating laws \eqref{18}, \eqref{19}, \eqref{23}, and \eqref{33}. Then, all the signals in the closed-loop system are bounded. Moreover, the tracking errors $z_{i,1}$, $i=1,\ldots,N$ tend toward a small neighborhood around zero.

\textbf{\emph{Proof:}} Define the following Lyapunov function
\begin{equation}
V=\sum\limits_{i = 1}^N {{V_{i,n_i}}}.\label{35}
\end{equation}

Invoking \eqref{34}, the $\alpha$th-order derivative of $V$ is denoted as
\begin{align}
{{\rm{{\cal D}}}^\alpha}V&=\sum\limits_{i = 1}^N {{{{\rm{{\cal D}}}^\alpha}V_{i,n_i}}}\leq\sum\limits_{i = 1}^N (-\lambda_{i,n_i}V_{i,n_i}+\zeta_{i,n_i})\nonumber\\
&=\sum\limits_{i = 1}^N\Big[-\lambda_{i,n_i}V_{i,n_i}+\sum\limits_{j = 1}^{n_i} \big(g_{i,j}\mathcal{N}(\delta_{i,j})\dot\delta_{i,j}+\dot\delta_{i,n_i}\big)\nonumber\\
&\ \ \ \ +\sum\limits_{j = 1}^{n_i}\frac{1}{4k_{i,j}}\bar\varepsilon_{i,j}^2 +\sum\limits_{j = 1}^{n_i}\frac{\rho_{i,j}}{2}\theta_{i,j}^{\ast T}\Lambda_{i,j}^{-1}\theta_{i,j}^\ast+\frac{1}{4b_i}\bar {d}_i^2\nonumber\\
&\ \ \ \ +\sum\limits_{j = 1}^{n_i}\frac{1}{{4{l_{i,j}}}}f_{i,j}^2-\mu_i z_{i,1}h_i+\frac{\gamma_{i,2}}{2\gamma_{i,1}}\mu_{i}^2\Big].\label{36}
\end{align}

According to Assumption 2, we obtain
\begin{equation}
\sum\limits_{j = 1}^{n_i}\frac{1}{{4{l_{i,j}}}}f_{i,j}^2\leq\sum\limits_{j = 1}^{n_i}\sum\limits_{q=1}^N\frac{1}{{4{l_{i,j}}}}\beta_{i,j,q}^2\psi_{i,j,q}^2(y_{q}).\label{37}
\end{equation}

Note that there is an unknown constant $\mu_i$, which satisfies $\mu_i\geq \sum\limits_{j = 1}^{n_i}\sum\limits_{q=1}^N\frac{1}{{4{l_{q,j}}}}\beta_{q,j,i}^2$. Hence, one obtains

\begin{equation}
\begin{aligned}
\sum\limits_{i = 1}^N&\Big[\sum\limits_{j = 1}^{n_i}\frac{1}{{4{l_{i,j}}}}f_{i,j}^2-\mu_i z_{i,1}h_i\Big]\\
&\leq\sum\limits_{i = 1}^N\Big[\sum\limits_{j = 1}^{n_i}\sum\limits_{q=1}^N\frac{1}{{4{l_{q,j}}}}\beta_{q,j,i}^2\psi_{q,j,i}^2(y_{q})-\mu_i z_{i,1}h_i\Big]\\
&\leq\sum\limits_{i = 1}^N\mu_i\frac{\varpi_i-z_{i,1}^2}{z_{i,1}^2+\varpi_i}\sum\limits_{j = 1}^{n_i}\sum\limits_{q=1}^N\psi_{q,j,i}^2(y_{q})=\sum\limits_{i = 1}^N\Psi_i.\label{38}
\end{aligned}
\end{equation}

Obviously, for $i=1,\ldots,N$, if $z_{i,1}^2>\varpi_i$, $\Psi_i<0$; instead, if $z_{i,1}^2\leq\varpi_i$, $z_{i,1}$ is bounded and $\bar\Psi_i$ exists such that $|\Psi_i|\leq\bar\Psi_i$, which is only related to the design parameter $\varpi_i$. With this relationship in mind, \eqref{36} is denoted as
\begin{align}
{{\rm{{\cal D}}}^\alpha}V\leq & -\sum\limits_{i = 1}^N\lambda_{i,n_i}V_{i,n_i}+\sum\limits_{i = 1}^N\sum\limits_{j = 1}^{n_i}(g_{i,j}\mathcal{N}(\delta_{i,j})\dot\delta_{i,j}+\dot\delta_{i,n_i})\nonumber\\
&  +\sum\limits_{i = 1}^N\Big[\sum\limits_{j = 1}^{n_i}(\frac{\rho_{i,j}}{2}\theta_{i,j}^{\ast T}\Lambda_{i,j}^{-1}\theta_{i,j}^\ast+\frac{1}{4k_{i,j}}\bar\varepsilon_{i,j}^2)\nonumber\\
&  +\frac{1}{4b_i}\bar {d}_i^2+\frac{\gamma_{i,2}}{2\gamma_{i,1}}\mu_{i}^2+\bar\Psi_i\Big]\nonumber\\
\leq & -\lambda V+\sum\limits_{i = 1}^N\sum\limits_{j = 1}^{n_i}\big(g_{i,j}\mathcal{N}(\delta_{i,j})\dot\delta_{i,j}+\dot\delta_{i,j}\big)+\zeta,\label{39}
\end{align}
where $\lambda=\min\{\lambda_{i,n_i},i=1,\ldots,N\}$ and $\zeta=+\sum\limits_{i = 1}^N\Big[\sum\limits_{j = 1}^{n_i}(\frac{\rho_{i,j}}{2}\theta_{i,j}^{\ast T}\Lambda_{i,j}^{-1}\theta_{i,j}^\ast+\frac{1}{4k_{i,j}}\bar\varepsilon_{i,j}^2)+\frac{1}{4b_i}\bar {d}_i^2+\frac{\gamma_{i,2}}{2\gamma_{i,1}}\mu_{i}^2+\bar\Psi_i\Big]$ are two positive constants.

According to Theorem 1, it is obvious that $V$ is bounded. Therefore, all signals of the closed-loop system remain bounded.

\textbf{\emph{Remark 6:}}
The work in \cite{u4,u8} only considered the control problem for integer-order systems with unknown control directions. By taking the integer-order integral for the inequality  $\dot V\leq-\lambda V+\sum\limits_{i = 1}^{n}\big(g_{i}\mathcal{N}(\delta_{i})\dot\delta_{i}+\dot\delta_{i}\big)+\zeta$, the exponential stability of $V$ can be established.  However, it is not applicable when the order is a non-integer. Therefore, we develop a new fractional-order Nussbaum function technique. From the inequality \eqref{39} of ${{\rm{{\cal D}}}^\alpha}V$, Theorem 1 can be used to show the stability of fractional-order interconnected systems. Meanwhile, in contrast with \cite{f10,f11} where the control directions are known, the extra term $\sum\limits_{i = 1}^N\sum\limits_{j = 1}^{n_i}\big(g_{i,j}\mathcal{N}(\delta_{i,j})\dot\delta_{i,j}+\dot\delta_{i,n_i}\big)$ can cause  many difficulties.

\section{Simulation}
Two examples will be provided in this section to verify the proposed results.
\subsection{Numerical example}
The considered fractional-order interconnected system composed of four subsystems is described as
\begin{equation}\label{42}
\begin{aligned}
&\left\{ {\begin{array}{*{20}{c}}
\begin{aligned}
&{{\rm{{\cal D}}}^\alpha}x_{i1}=x_{i2}+\phi_{i1}\\
&{{\rm{{\cal D}}}^\alpha}x_{i2}=g_iu_i+\phi_{i2}+d_i+f_{i2}\\
&y_i=x_{i1},\ i=1,\ldots,4,\\
\end{aligned}
\end{array}} \right.\\
\end{aligned}
\end{equation}
where $\alpha=0.8$, the completely unknown control gains $g_1=2+\sin t$, $g_2=2$, $g_3=3-\cos t$, $g_4=3$, the unknown functions $\phi_{11}=0.6x_{11}^2$, $\phi_{12}=\frac{x_{12}}{1+x_{11}^2}$, $\phi_{21}=0.5e^{-x_{21}^2}$, $\phi_{22}=e^{-x_{22}^2}\sin{x_{21}}$, $\phi_{31}=x_{31}^2$, $\phi_{32}=(1-x_{31}^2)x_{12}$, $\phi_{41}=0$, $\phi_{42}=x_{41}x_{42}^2$, the external disturbances $d_1=d_2=0.3\cos(\pi t)$, $d_3=d_4=0.4\sin(\pi t)$, the unknown interactions $f_{12}=0.5x_{21}+x_{31}+\sin x_{41}$, $f_{22}=x_{11}+0.6x_{31}+0.7x_{41}$, $f_{32}=x_{11}+\sin x_{21}+x_{41}$,  $f_{42}=x_{11}+\sin x_{21}+0.6x_{31}$.
The initial condition is $x_i(0)=[0.1,0.1]^T$ for $i=1,\ldots,4$. Let the reference signals be $y_{ri}=\sin2t$, $i=1,\ldots,4$. The Nussbaum functions are chosen as $N(\delta)=\delta^2\sin{\frac{\delta}{2}\pi}$, which can also be selected as other common forms, such as $N(\delta)=e^{\delta^2}\cos{\frac{\delta}{2}\pi}$, $N(\delta)=e^{\delta^2}\sin{\frac{\delta}{2}\pi}$, and $N(\delta)=\delta^2\cos{\frac{\delta}{2}\pi}$.

%
%
%

The design parameters are selected as $\bar c_{11}=3$, $\bar c_{21}=5$, $\bar c_{31}=4$, $\bar c_{41}=3$, $\bar c_{i2}=1$, $\Lambda_{i,j}=1$, and $\gamma_{i,1}=1$, where $\bar c_{i1}=c_{i1}+k_{i1}+l_{i1}$ and $\bar c_{i2}=c_{i2}+k_{i2}+l_{i2}+b_i$, $i=1,\ldots,4$, $j=1,2$. In the FLSs, we use the Gaussian membership functions $\mu_{F_{i,j}^l}(x_{ij})=e^{[-0.5(x_{ij}-3+l)^2]}$, $l=1,\ldots,5$ to determine the fuzzy basis $\varphi_{ij}(\bar x_{ij})$. Let the initial value of $\theta_{i,j}$ be zero.

The simulation results are illustrated in Figs.  \ref{fig1}--\ref{fig4}. It is observed that, in Fig. \ref{fig1}, the outputs $y_i$, $i=1,\ldots,4$ can track the reference signals well with small tracking errors. Meanwhile, the trajectories of tracking errors $z_{i1}$ for $i=1,\ldots,4$ are shown in Fig. \ref{fig2}, which converge to the neighborhood of zero. The variations of the control input and parameters of FLSs are plotted in Figs. \ref{fig3}--\ref{fig4}. We can see that these signals remain bounded.

\subsection{Practical example}
According to \cite{motor,motor_f}, we consider a fractional-order smooth-air-gap permanent magnet synchronous motor (PMSM) as follows:
\begin{equation}
\begin{aligned}
{{\rm{{\cal D}}}^\alpha}\omega&=\kappa(i_q-\omega),\\
{{\rm{{\cal D}}}^\alpha}i_q&=-i_q-\omega i_d+\nu\omega+g_1u_q,\\
{{\rm{{\cal D}}}^\alpha}i_d&=-i_d+\omega i_q+g_2u_d,\\
\end{aligned}\label{41}
\end{equation}
where $\alpha=0.9$, $\omega$, $i_q$ and $i_d$ denote rotor angular velocity and $d-q$ axis currents and the parameters $\kappa$, $\nu$ and $g_i$, $i=1,2$ are determined by the specifications of PMSM. It is demonstrated in many references \cite{fa4,motor_f} that PMSM can be better described  by fractional-order models.
It is noted that PMSM is an MIMO system such that $u=[u_q,u_d]^T$ and $y=[\omega,i_d]$. Obviously, coupling exists  in the systems \eqref{41}. Therefore, the PMSM can be regarded as an interconnected system and divided into two subsystems:
\begin{equation}\label{42}
\begin{aligned}
{{\rm{{\cal D}}}^\alpha}x_{11}&=\kappa(x_{12}-x_{11}),\\
{{\rm{{\cal D}}}^\alpha}x_{12}&=-x_{12}-x_{11} x_{21}+\nu x_{11}+g_1u_q,\\
y_1&=x_{11},\\
{{\rm{{\cal D}}}^\alpha}x_{21}&=-x_{21}+x_{11} x_{12}+g_2u_d,\\
y_2&=x_{21},\\
\end{aligned}
\end{equation}
where $x_1=[x_{11},x_{12}]^T=[\omega,i_q]^T$, $x_{21}=i_d$. $\kappa$, $g_1$ and $g_2$ are unknown control coefficients with unknown signs. We set the parameters as $\kappa=2$, $\nu=3$, and $g_1=g_2=3$. The initial condition is selected as $[x_{11},x_{12},x_{21}]=[0.1,0.1,0.1]$. The reference signals are $y_{r1}=\sin2t$ and $y_{r2}=0$. The design parameters are chosen as $\bar c_{11}=c_{11}+k_{11}=10$, $\bar c_{12}=c_{12}+k_{12}+l_{12}+b_{12}=3$, $\bar c_{21}=c_{21}+k_{21}+l_{21}+b_{21}=3$, $\Lambda_{11}=\Lambda_{12}=\Lambda_{21}=1$, and $\gamma_{11}=\gamma_{21}=1$. To compute the fuzzy basis $\varphi_{ij}(\bar x_{ij})$, the fuzzy membership functions in FLSs are chosen as $\mu_{F_{i,j}^l}(x_{ij})=e^{[-0.5(x_{ij}-3+l)^2]}$, $l=1,\ldots,5$. The initial value of $\theta_{i,j}$ is set as zero.

The corresponding simulation results are exhibited in Figs. \ref{fig5}--\ref{fig8}, whose qualitative analysis is similar to that in Example A. It is clearly seen that the proposed control method is effective for improving the dynamic behavior for the PMSM system (i.e., it achieves a good tracking performance).
%

To show the robustness of our proposed method, we conduct two comparative simulations by reasonably changing the parameters of the systems \eqref{42} shown above under the same design parameters. First, we present Fig. \ref{fig9}, which simultaneously depicts the trajectories of tracking errors in two sets of parameters $\kappa$ and $\nu$. It can be seen that, when the parameters of PMSM are changed, the tracking performance almost remains the same. Second, we change the fractional orders that affect the dynamic characteristics of PMSM \cite{motor_f}. The tracking errors are shown in Fig. \ref{fig10}. It is observed that a good tracking performance can be obtained. The above two comparisons demonstrate the robustness of our proposed method.

\section{Conclusion}
In this paper, we propose a new fractional-order Nussbaum function technique  for unknown control directions in fractional-order systems. With the help of this technique, a decentralized adaptive fuzzy control method is developed for a class of interconnected systems with unknown identical control directions. In future work, more complex fractional-order interconnected systems will be investigated.

\appendices
\section{Proof of the Theorem 1}
The proof is completed in two parts.

\textbf{Part 1.} Let us focus on the bound of $V$. From the inequality \eqref{6}, we obtain
\begin{equation}
{{\rm{{\cal D}}}^\alpha}V+M(t)= -\lambda V+\sum\limits_{i = 1}^n\big(g_i\mathcal{N}(\delta_i)\dot\delta_i+\dot\delta_i\big)+\zeta,\label{43}
\end{equation}
where $M(t)$ is the nonnegative time-varying function.
Through the Laplace transform, we obtain
\begin{equation}
\begin{aligned}
V(s)=&\frac{s^{\alpha-1}}{s^\alpha+\lambda}V(0)-\frac{1}{s^\alpha+\lambda}M(s)\\
&+\frac{1}{s^\alpha+\lambda}f(s)+\frac{\zeta}{s(s^\alpha+\lambda)},\label{44}
\end{aligned}
\end{equation}
where $f(t)=\sum\limits_{i = 1}^n\big(g_i\mathcal{N}(\delta_i)\dot\delta_i+\dot\delta_i\big)$. Then, taking the Laplace inverse transform, we obtain
\begin{align}
V(t)=&E_{\alpha,1}(-\lambda t^\alpha)V(0)-M(t)\ast \big(t^{\alpha-1}E_{\alpha,\alpha}(-\lambda t^\alpha)\big)\nonumber \\
&+ f(t)\ast \big(t^{\alpha-1}E_{\alpha,\alpha}(-\lambda t^\alpha)\big)\nonumber\\
&+\zeta t^\alpha E_{\alpha,\alpha+1}(-\lambda t^\alpha),\label{45}
\end{align}
where $\ast$ is the convolution operator.

Note that $t^{\alpha-1}E_{\alpha,\alpha}(-\lambda t^\alpha)\geq 0$. Therefore, we can conclude that $M(t)\ast \big(t^{\alpha-1}E_{\alpha,\alpha}(-\lambda t^\alpha)\big)\geq 0$. According to Lemma 3, we obtain
\begin{equation}
|\zeta t^\alpha E_{\alpha,\alpha+1}(-\lambda t^\alpha)|\leq \frac{\zeta t^\alpha \sigma}{1+|\lambda t^\alpha|}\leq\frac{\zeta \sigma}{\lambda}.\label{46}
\end{equation}

Furthermore, according to  Lemma 1, it can be found that $E_{\alpha,1}(-\lambda t^\alpha)V(0)$ is bounded and $\lim\limits_{t\rightarrow \infty}E_{\alpha,1}(-\lambda t^\alpha)V(0)=0$. Now, let us analyze  $f(t)\ast \big(t^{\alpha-1}E_{\alpha,\alpha}(-\lambda t^\alpha)\big)$, which is expressed as
\begin{equation}
\begin{aligned}
f(t)\ast &\big(t^{\alpha-1}E_{\alpha,\alpha}(-kt^\alpha)\big)\\
&=\int\limits_{-\infty}^\infty\Big(\sum\limits_{i = 1}^ng_i\mathcal{N}(\delta_i)\dot\delta_i+\dot\delta_i\Big)(t-\varsigma)^{\alpha-1}\\
&\ \ \ \ \times E_{\alpha,\alpha}(-\lambda(t-\varsigma)^\alpha)d\varsigma.\label{47}
\end{aligned}
\end{equation}

Due to the fact that $f(t)$ and $t^{\alpha-1}E_{\alpha,\alpha}(-\lambda t^\alpha)$ are defined on $[0,\infty)$, \eqref{47} is rewritten as
\begin{align}
&\int\limits_0^t\Big(\sum\limits_{i = 1}^ng_i\mathcal{N}(\delta_i)\dot\delta_i+\dot\delta_i\Big)E_{\alpha,\alpha}\big(-\lambda (t-\varsigma)^\alpha\big)(t-\varsigma)^{\alpha-1}d\varsigma\nonumber\\
&=\sum\limits_{i = 1}^n\int\limits_0^tg_i\mathcal{N}(\delta_i)\dot\delta_iE_{\alpha,\alpha}\big(-\lambda(t-\varsigma)^\alpha\big)(t-\varsigma)^{\alpha-1}d\varsigma\nonumber\\
&\ \ \ \ +\sum\limits_{i = 1}^n\int\limits_0^t\dot\delta_iE_{\alpha,\alpha}\big(-\lambda(t-\varsigma)^\alpha\big)(t-\varsigma)^{\alpha-1}d\varsigma.\label{48}
\end{align}

Based on the above analysis, \eqref{45} becomes
\begin{align}
V(t)\leq &\sum\limits_{i = 1}^n\int\limits_0^tg_i\mathcal{N}(\delta_i)\dot\delta_iE_{\alpha,\alpha}\big(-\lambda(t-\varsigma)^\alpha\big)(t-\varsigma)^{\alpha-1}d\varsigma\nonumber\\
&+ \sum\limits_{i = 1}^n\int\limits_0^t\dot\delta_iE_{\alpha,\alpha}\big(-\lambda(t-\varsigma)^\alpha\big)(t-\varsigma)^{\alpha-1}d\varsigma\nonumber\\
&+H,\label{49}
\end{align}
where $H= E_{\alpha,1}(-\lambda t^\alpha)V(0)+\frac{\zeta \sigma}{\lambda}$ is bounded.

\textbf{Part 2.} We analyze the boundedness of $V$ from the inequality \eqref{49} based on the properties of the Nussbaum function. For convenience, define $V_g(t_p,t_q)$ as
\begin{align}
V_g(t_p&,t_q)=V_g\big(\delta(t_p),\delta(t_q)\big)=V_g(\delta_p,\delta_q)\nonumber\\
=&\sum\limits_{i = 1}^n\int\limits_{t_p}^{t_q}g_i\mathcal{N}(\delta_i)\dot\delta_iE_{\alpha,\alpha}(-\lambda(t-\varsigma)^\alpha)(t-\varsigma)^{\alpha-1}d\varsigma\nonumber\\
&+\sum\limits_{i = 1}^n\int\limits_{t_p}^{t_q}\dot\delta_iE_{\alpha,\alpha}(-\lambda(t-\varsigma)^\alpha)(t-\varsigma)^{\alpha-1}d\varsigma,\label{50}
\end{align}
and   $V_{gi}(t_p,t_q)$ for $i=1,\ldots,n$ as
\begin{equation}
\begin{aligned}
V_{gi}(t_p,t_q)=&V_{gi}(\delta(t_p),\delta(t_q))=V_{gi}(\delta_p,\delta_q)\\
=&\int\limits_{t_p}^{t_q}(g_i\mathcal{N}(\delta_i)\dot\delta_i+\dot\delta_i)E_{\alpha,\alpha}(-\lambda(t-\varsigma)^\alpha)\\
&\times(t-\varsigma)^{\alpha-1}d\varsigma,\label{51}
\end{aligned}
\end{equation}
where $t_p\leq t_q$. To accomplish the proof, we select the Nussbaum function as $\mathcal{N}(\delta)=e^{\delta^2}\sin{\frac{\delta}{2}\pi}$. It is obvious that $\mathcal{N}(\delta)$ is positive for $\delta\in (\delta_{m1},\delta_{m2})=(4m,4m+2)$  and  negative for $\delta\in (\delta_{m2},\delta_{m3})=(4m+2,4m+4)$ with $m\in\mathbb{Z}$.

From the above definitions, $V_g(t_0,t)$ for $t\in[t_0,t_f)$ can be rewritten as
\begin{align}
V_g&(t_0,t)\nonumber\\
&=\sum\limits_{i = 1}^n\int\limits_{t_0}^{t}g_i\mathcal{N}(\delta_i)\dot\delta_iE_{\alpha,\alpha}(-\lambda(t-\varsigma)^\alpha)(t-\varsigma)^{\alpha-1}d\varsigma\nonumber\\
&\ \ \ \ +\sum\limits_{i = 1}^n\int\limits_{t_0}^{t}\dot\delta_iE_{\alpha,\alpha}(-\lambda(t-\varsigma)^\alpha)(t-\varsigma)^{\alpha-1}d\varsigma\nonumber\\
&=\sum\limits_{i = 1}^\omega V_{gi}(t_0,t)+\sum\limits_{i = \omega+1}^nV_{gi}(t_0,t).\label{52}
\end{align}

In the following, we will show that $\delta_i$ is bounded by seeking a contradiction. We suppose that $\delta_i$ is unbounded. In particular, $\delta_1,\ldots,\delta_\omega$ are unbounded, but $\delta_{\omega+1},\ldots,\delta_n$ are bounded for $1\leq\omega\leq n$. Therefore,  three cases  should be considered: 1) $\delta_i$ has no upper bound; 2) $\delta_i$ has no lower bound, and 3) $\delta_1,\ldots,\delta_j$ have no upper bound, while $\delta_{j+1},\ldots,\delta_\omega$ have no lower bound, $i=1,\ldots,\omega$ and $j=1,\ldots,\omega-1$.

\emph{Case 1}: $\delta_i(1\leq i\leq\omega)$ has no upper bound on $[0,t_f)$.

In this case, there exist a monotonically increasing variable $\delta_i=\delta_i(t_m)$, $\lim\limits_{m\rightarrow \infty}t_m\rightarrow t_f$, and $\lim\limits_{t\rightarrow t_f}\delta_i\rightarrow \infty$. Moreover, there exist time instants $t_{i1}$ and $t_{i2}$ which are defined as $t_{i1}=\{t:\delta_i=\delta_{m1}\}$ and $t_{i2}=\{t:\delta_i=\delta_{m2}\}$. From  \eqref{52}, $V_g(t_0,t)$ is divided into two sums over $[t_0,t_f)$ as
\begin{equation}
V_g(t_0,t_f)=\sum\limits_{i = 1}^\omega V_{gi}(t_0,t_f)+\sum\limits_{i = \omega+1}^nV_{gi}(t_0,t_f).\label{53}
\end{equation}

(i) In the first sum of \eqref{53}, the $\delta_i$ of $V_{gi}$ is unbounded for $1\leq i\leq\omega$. Define $g_{\max}=\max_{1\leq i\leq \omega}\{|g_i(t)|\}$ and $g_{\min}=\min_{1\leq i\leq \omega}\{|g_i(t)|\}$.

Firstly, let us consider $V_{gi}(1\leq i\leq\omega)$ with $g_i(t)<0$.

By defining $F_1=\inf_{x\in[w^-, w^+]}F(x)$ and $F_2=\sup_{x\in[w^-, w^+]}F(x)$, the integral inequality is shown as $(w^+-w^-)F_1\leq \int_{w^-}^{w^+} F(x)dx\leq (w^+-w^-) F_2$. Noting that $0< E_{\alpha,\alpha}(-\lambda(t-\varsigma)^\alpha)(t-\varsigma)^{\alpha-1}\leq1$ with $0<\alpha<1$ for $\varsigma\in[t_0,t_{i1}]$, according to the integral inequality, we obtain
\begin{equation}
\begin{aligned}
|V_{gi}(\delta_0,\delta_{m1})|&\leq(\delta_{m1}-\delta_0)g_{\max}e^{(4m)^2}(t_f-t_{i1})^{\alpha-1}\\
&\ \ \ \ +(\delta_{m1}-\delta_0)(t_f-t_{i1})^{\alpha-1}\\
&=(4m-\delta_{0})g_{\max}e^{(4m)^2}(t_f-t_{i1})^{\alpha-1}\\
&\ \ \ \ +(4m-\delta_{0})(t_f-t_{i1})^{\alpha-1}.\label{54}
\end{aligned}
\end{equation}

Due to $\mathcal{N}(\delta_i)\geq0$ for $\delta_i\in [\delta_{m1},\delta_{m2}]$, we obtain
\begin{align}
&V_{gi}(\delta_{m1},\delta_{m2})\nonumber \\
&\leq\int\limits_{4m+1-c_{mi}}^{4m+1+c_{mi}}g_i\mathcal{N}(\delta_i)E_{\alpha,\alpha}(-\lambda(t_f-\varsigma)^\alpha)(t_f-\varsigma)^{\alpha-1}d\delta_i\nonumber\\
&\ \ \ \ +\int\limits_{4m+1-c_{mi}}^{4m+1+c_{mi}}E_{\alpha,\alpha}(-\lambda(t_f-\varsigma)^\alpha)(t_f-\varsigma)^{\alpha-1}d\delta_i\nonumber\\
&\leq -2c_{mi}g_{\min}\inf\limits_{\delta_i\in [\delta_{m1},\delta_{m2}]}{\mathcal{N}(\delta_i)}\nonumber\\
&\ \ \ \ \times E_{\alpha,\alpha}(-\lambda(t_f-t_{i1})^\alpha)(t_f-t_{i1})^{\alpha-1}\nonumber\\
&\ \ \ \ +2c_{mi}E_{\alpha,\alpha}(-\lambda(t_f-t_{i2})^\alpha)(t_f-t_{i2})^{\alpha-1}\nonumber\\
&=\big(-a_{i1}e^{(4m+1-c_{mi})^2}+a_{i2}\big)(t_f-t_{i1})^{\alpha-1},\label{55}
\end{align}
where $a_{i1}=2c_{mi}g_{\min}\cos{(\frac{c_{mi}}{2}\pi)}E_{\alpha,\alpha}(-\lambda(t_f-t_{i1})^\alpha)>0$, $a_{i2}=2c_{mi}E_{\alpha,\alpha}(-\lambda(t_f-t_{i1})^\alpha)$, and $c_{mi}\in(0,1)$. Hence, it follows from \eqref{54} and \eqref{55} that
\begin{align}
V_{gi}(\delta_0,\delta_{m2})&=V_{gi}(\delta_0,\delta_{m1})+V_{gi}(\delta_{m1},\delta_{m2})\nonumber\\
&\leq(4m-\delta_{0})g_{\max}e^{(4m)^2}(t_f-t_{i1})^{\alpha-1}\nonumber\\
&\ \ \ \ +(4m-\delta_{0})(t_f-t_{i1})^{\alpha-1}\nonumber\\
&\ \ \ \ +(-a_{i1}e^{(4m+1-c_{mi})^2}+a_{i2})(t_f-t_{i1})^{\alpha-1}\nonumber\\
&=e^{(4m)^2}\big(-a_{i1}e^{[8m(1-c_{mi})+(1-c_{mi})^2]}\nonumber\\
&\ \ \ \ +(4m-\delta_0)g_{\max}+\frac{4m-\delta_0+a_{i2}}{e^{(4m)^2}}\big)\nonumber\\
&\ \ \ \ \times(t_f-t_{i1})^{\alpha-1}.\label{56}
\end{align}

Note that $(t_f-t_{i1})^{\alpha-1} >0$, $1-c_{mi}>0$ and $e^m$ grow   much faster than $m$. Therefore, we find that $V_{gi}(\delta_0,\delta_{m2})=V_{gi}(\delta_0,4m+2)\rightarrow-\infty$ when $m \rightarrow+\infty$, which means that $V_{gi}(t_0,t_f)\rightarrow-\infty$.

Secondly, we analyze $V_{gi}(1\leq i\leq\omega)$ with $g_i(t)>0$. Similar to the above derivation, using the integral inequality in the interval $[\delta_0,\delta_{m2}]$, we obtain
\begin{align}
|V_{gi}(\delta_0,\delta_{m2})|&\leq(\delta_{m2}-\delta_0)g_{\max}e^{(4m+2)^2}(t_f-t_{i2})^{\alpha-1}\nonumber\\
&\ \ \ \ +(\delta_{m2}-\delta_0)(t_f-t_{i2})^{\alpha-1}\nonumber\\
&=(4m+2-\delta_{0})g_{\max}e^{(4m+2)^2}(t_f-t_{i2})^{\alpha-1}\nonumber\\
&\ \ \ \ +(4m+2-\delta_{0})(t_f-t_{i2})^{\alpha-1}.\label{57}
\end{align}

It is known that $\mathcal{N}(\delta_i)\leq 0$ when $\delta_i\in[\delta_2,\delta_3]$. Thus, we obtain
\begin{align}
&V_{gi}(\delta_{m2},\delta_{m3}) \nonumber\\
&\leq\int\limits_{4m+3-c_{mi}}^{4m+3+c_{mi}}g_i\mathcal{N}(\delta_i)E_{\alpha,\alpha}(-\lambda(t_f-\varsigma)^\alpha)(t_f-\varsigma)^{\alpha-1}d\delta_i\nonumber\\
&\ \ \ \ +\int\limits_{4m+3-c_{mi}}^{4m+3+c_{mi}}E_{\alpha,\alpha}(-\lambda(t_f-\varsigma)^\alpha)(t_f-\varsigma)^{\alpha-1}d\delta_i\nonumber\\
&\leq2c_{mi}g_{\min}\sup{\mathcal{N}(\delta_i)}E_{\alpha,\alpha}(-\lambda(t_f-t_{i2})^\alpha)(t_f-t_{i2})^{\alpha-1}\nonumber\nonumber\\
&\ \ \ \ +2c_{mi}E_{\alpha,\alpha}(-\lambda(t_f-t_{i2})^\alpha)(t_f-t_{i2})^{\alpha-1}\nonumber\\
&=(-a_{i1}e^{(4m+3-c_{mi})^2}+a_{i2})(t_f-t_{i2})^{\alpha-1},\label{58}
\end{align}
where $a_{i1}=2c_{mi}g_{\min}\cos{(\frac{c_{mi}}{2}\pi)}E_{\alpha,\alpha}(-\lambda(t_f-t_{i2})^\alpha)>0$ and $a_{i2}=2c_{mi}E_{\alpha,\alpha}(-\lambda(t_f-t_{i2})^\alpha)$. Hence, from \eqref{57} and \eqref{58}, we obtain
\begin{align}
V_{gi}(\delta_0,\delta_{m3})&=V_{g-\i}(\delta_0,\delta_{m2})+V_{g-\i}(\delta_{m2},\delta_{m3})\nonumber\\
&\leq(4m+2-\delta_0)g_{\max}e^{(4m+2)^2}(t_f-t_{i2})^{\alpha-1}\nonumber\\
&\ \ \ \ +(4m+2-\delta_0)(t_f-t_{i2})^{\alpha-1}\nonumber\\
&\ \ \ \ +(-a_{i1}e^{(4m+3-c_{mi})^2}+a_{i2})(t_f-t_{i2})^{\alpha-1}\nonumber\\
&=e^{(4m+2)^2}\big(-a_{i1}e^{[4(2m+1)(1-c_{mi})+(1-c_{mi})^2]}\nonumber\\
&\ \ \ \ +(4m+2-\delta_0)g_{\max}\nonumber\\
&\ \ \ \ +\frac{4m+2-\delta_0+a_{i2}}{e^{(4m)^2}}\big)(t_f-t_{i2})^{\alpha-1}.\label{59}
\end{align}

Similar to \eqref{56}, we know that $V_{gi}(\delta_0,\delta_{m3})=v_{gi}(\delta_0,4m+4)\rightarrow-\infty$ when $m \rightarrow+\infty$, which means that $V_{gi}(t_0,t_f)\rightarrow-\infty$.
Hence, a subsequence that results in $V_{gi}(t_0,t_f)\rightarrow-\infty$ for $1\leq i\leq\omega$ can always be found whether $g_i(t)$ is positive or negative.

Suppose that $\delta_{\bar q}$, $\bar q\in[1,\omega]$ grows the fastest. Therefore, for $g_{i}(t)<0$, when $\delta_{\bar q}=\delta_{m2}$, i.e., $t_m=t_{\bar q2}$, from \eqref{54} and \eqref{56}, we obtain
\begin{align}
\sum\limits_{i = 1}^\omega V_{gi}(t_0,t_{\bar q2})&\leq V_{g\bar q}(\delta_0,\delta_{m2})+\sum\limits_{i = 1,i\neq\bar q}^\omega|V_{gi}(\delta_0,\delta_{m1})|\nonumber\\
& \leq e^{(4m)^2}\big(-a_{\bar q21}e^{[8m(1-c_{m\bar q2})+(1-c_{m\bar q2})^2]}\nonumber\\
&\ \ \ \ +\bar\omega(4m-\delta_0)g_{\max}+\frac{\bar\omega(4m-\delta_0)+a_{i2}}{e^{(4m)^2}}\big)\nonumber\\
&\ \ \ \ \times(t_f-t_{i1})^{\alpha-1},
\end{align}
where $\bar \omega={\omega(t_f-t_{i2})^{\alpha-1}}/{(t_f-t_{i1})^{\alpha-1}}$ is bounded  since $\delta_i$ is monotonically increasing.

According to the conclusion of \eqref{56}, $\sum\limits_{i = 1}^\omega V_{gi}(t_0,t_{\bar q2})\rightarrow-\infty$ when $t_{\bar q2}\rightarrow t_f$; that is, $\sum\limits_{i = 1}^\omega V_{gi}(t_0,t_f)\rightarrow-\infty$. Through the similar analysis for $g_{i}(t)>0$, we find that $\sum\limits_{i = 1}^\omega V_{gi}(t_0,t_f)\rightarrow-\infty$.
Consequently, we can obtain   $\sum\limits_{i = 1}^\omega V_{gi}(t_0,t_f)\rightarrow-\infty$.

(ii) In the second sum of \eqref{53}, $\delta_i$ from $V_{gi}$ is bounded for $\omega+1\leq i\leq n$. With the time interval $[t_0,t_f)$ divided into $[t_0,t_f-1]$ and $(t_f-1,f_f)$, $V_{gi}$ is rewritten as
\begin{equation}
\begin{aligned}
V_{gi}&(t_0,t_f)=V_{gi}(t_0,t_f-1)+V_{gi}(t_f-1,t_f)\\
&=\int\limits_{t_0}^{t_f-1}\big(g_i\mathcal{N}(\delta_i)\dot\delta_i+\dot\delta_i\big)E_{\alpha,\alpha}(-\lambda(t_f-\varsigma)^\alpha)\\
&\ \ \ \ \times (t_f-\varsigma)^{\alpha-1}d\tau +\int\limits_{t_f-1}^{t_f}\big(g_iN(\delta_i)\dot\delta_i+\dot\delta_i\big)\\
&\ \ \ \ \times E_{\alpha,\alpha}(-\lambda(t_f-\varsigma)^\alpha)(t_f-\varsigma)^{\alpha-1}d\varsigma.\label{60}
\end{aligned}
\end{equation}

For $\varsigma\in[t_0,t_f-1]$, it is obvious that $E_{\alpha,\alpha}(-\lambda(t_f-\varsigma)^\alpha)<0$ and $(t_f-\varsigma)^{\alpha-1}\leq0$ with $\alpha\in(0,1)$. Using the integral inequality, the term $V_{gi}(t_o,t_f-1)$ in \eqref{60} is denoted as
\begin{align}
\int\limits_{t_0}^{t_f-1}&\big(g_iN(\delta_i)\dot\delta_i+\dot\delta_i\big)E_{\alpha,\alpha}(-\lambda(t_f-\varsigma)^\alpha)(t_f-\varsigma)^{\alpha-1}d\varsigma\nonumber\\
&\leq(\delta_{i\max}-\delta_{i\min})(g_{\max}e^{\delta_{i\max}^2}+1),\label{61}
\end{align}
where $\delta_{i\max}=\sup_{t\in[t_0,t_f]}\delta_i(t)$, $\delta_{i\min}=\inf_{t\in[t_0,t_f]}\delta_i(t)$.

For $\varsigma\in(t_f-1,t_f)$, it is known that $(t_f-\varsigma)^{\alpha-1}>1$ and $\lim\limits_{\varsigma \rightarrow t_f}(t_f-\varsigma)^{\alpha-1}\rightarrow\infty$, where the comparison test of the improper integral will be used to analyze  $V_{gi}(t_f-1,t_f)$. Since $\delta_i$ is bounded, $g_i\mathcal{N}(\delta_i)\dot\delta_i+\dot\delta_i$ is bounded. Hence, we obtain
\begin{align}
\int\limits_{t_f-1}^{t_f}&(g_i\mathcal{N}(\delta_i)\dot\delta_i+\dot\delta_i)E_{\alpha,\alpha}(-\lambda(t_f-\varsigma)^\alpha)(t_f-\varsigma)^{\alpha-1}d\varsigma\nonumber\\
&\leq\int\limits_{t_f-1}^{t_f}\kappa E_{\alpha,\alpha}(-\lambda(t_f-\varsigma)^\alpha)(t_f-\varsigma)^{\alpha-1}d\varsigma,\label{62}
\end{align}
where $\kappa$ is a positive constant and satisfies $|g_i\mathcal{N}(\delta_i)\dot\delta_i+\dot\delta_i|\leq \kappa$. At the present stage, we construct a limitation as
\begin{equation}
\lim\limits_{t\rightarrow t_f^-}(t_f-t)^{1-\alpha}\kappa(t_f-t)^{\alpha-1}E_{\alpha,\alpha}(-\lambda(t_f-t)^\alpha)=\kappa.\label{63}
\end{equation}

Due to $0\leq \kappa\leq\infty$ and $1-\alpha<1$, we directly conclude that $\int\limits_{t_f-1}^{t_f}\kappa E_{\alpha,\alpha}(-\lambda(t_f-\varsigma)^\alpha)(t_f-\varsigma)^{\alpha-1}d\varsigma$ is a convergence; that is, $V_{gi}(t_f-1,t_f)$ is bounded. As a result, it is derived that
\begin{equation}
V_{gi}(t_f-1,t_f)\leq \hbar,\label{64}
\end{equation}
where $\hbar$ is a positive constant. Combining \eqref{61} and \eqref{64}, we obtain
\begin{equation}
\begin{aligned}
V_{gi}(t_0,t_f)\leq (\delta_{i\max}-\delta_{i\min})(g_{\max}e^{\delta_{i\max}^2}+1)+\hbar.\label{65}
\end{aligned}
\end{equation}

Consequently, we can determine that $V_{gi}(t_0,t_f)(\omega+1\leq i \leq n)$ and $\sum\limits_{i = \omega+1}^nV_{gi}(t_0,t_f)$ are bounded.

Finally, according to the above analysis of $($i$)$ and $($ii$)$,  $V_g(t_0,t_f)$ from \eqref{53} satisfies the constraint that $V_g(t_0,t_f)\rightarrow-\infty$. From \eqref{49}, we further conclude that  $V(t)\rightarrow-\infty$ when $t\rightarrow t_f$. However, according to the definition of $V(t)$, $V(t)$ is nonnegative at any time $t$. Thus, a sequence that results in a contradiction can always be  found  in \eqref{49}. As a result, $\delta_i(1\leq i \leq \omega)$ has an upper bound on $[t_0,t_f)$.

\emph{Case 2}: $\delta_i(1\leq i \leq \omega)$ has no lower bound on $[t_0,f_f)$.

First, we define $\delta_i= -\xi_i$ for $i=1,2,\ldots,n$. Therefore, the condition of $\delta_i$ is transformed into  $\xi_i(1\leq i \leq \omega)$, which has no upper bound. Due to the Nussbaum function $\mathcal{N}(\cdot)$ in \eqref{5} being an odd function, we can rewrite \eqref{50} as
\begin{align}
V_g&(t_0,t)\nonumber\\
&=\sum\limits_{i = 1}^n\int\limits_{t_0}^tg_i\mathcal{N}(\xi_i)\dot \xi_iE_{\alpha,\alpha}(-\lambda(t-\varsigma)^\alpha)(t-\varsigma)^{\alpha-1}d\varsigma\nonumber\\
&\ \ \ \ -\sum\limits_{i = 1}^n\int\limits_{t_0}^t\dot \xi_iE_{\alpha,\alpha}(-\lambda(t-\varsigma)^\alpha)(t-\varsigma)^{\alpha-1}d\varsigma.\label{66}
\end{align}

Since   $\xi_i(1\leq i \leq \omega)$ has no upper bound, the proof is accomplished with similar procedures in \emph{Case 1}. Hence, $\xi_i(1\leq i \leq \omega)$ must have an upper bound by constructing a sequence to result in a contradiction. Based on the definition of $\xi_i(1\leq i \leq \omega)$, we can determine that $\delta_i(1\leq i \leq \omega)$ must have a lower bound on the time interval $[t_0,t_f)$.

\emph{Case 3}: $\delta_i(1\leq i \leq j)$ has no upper bound, while $\delta_i(j+1\leq i \leq \omega)$ has no lower bound for $j=1,2,\ldots,\omega-1$.

In this case, to differentiate from \emph{{Case} 1} and \emph{Case} \emph{2},  $\omega$ is supposed to satisfy $\omega\in[2,n]$. Obviously, when $n=1$, the multiple Nussbaum functions problem will be simplified to a single Nussbaum function problem. The proof can be provided by \emph{Cases 1} and \emph{2}. Thus, in the following, we will analyze \emph{Case 3} for $n\geq 2$.

The analysis can be established by combining the results in \emph{Cases 1} and \emph{2}. Firstly, we define $\delta_i=-\xi_i$ for $j+1\leq i \leq \omega$. Thus, the initial condition is transformed into that $\delta_i(1\leq i \leq j)$ and $\xi_i(j+1\leq i \leq \omega)$ have no upper bound. Hence, we can divide $(65)$ into three terms as
\begin{equation}
\begin{aligned}
V_g(t_0,t)=\bar V_{g1}(t_0,t)+ \bar V_{g2}(t_0,t)+\bar V_{g3}(t_0,t),\label{67}
\end{aligned}
\end{equation}
where $\bar V_{gi}(i=1,2,3)$ relate to the corresponding variables that have different bound in the initial condition.

In the $\bar V_{g1}(t_0,t)$, the variable $\delta_i(1\leq i \leq j)$ has no upper bound. Thus, $\bar V_{g1}$ is expressed as
\begin{align}
V_{g1}&(t_0,t)\nonumber\\
&=\sum\limits_{i = 1}^j\int\limits_{t_0}^tg_i\mathcal{N}(\delta_i)\dot\delta_iE_{\alpha,\alpha}(-\lambda(t-\varsigma)^\alpha)(t-\varsigma)^{\alpha-1}d\varsigma\nonumber\\
&\ \ \ \ +\sum\limits_{i = 1}^j\int\limits_{t_0}^t\dot\delta_iE_{\alpha,\alpha}(-\lambda(t-\varsigma)^\alpha)(t-\varsigma)^{\alpha-1}d\varsigma.\label{68}
\end{align}

The boundedness of $\bar V_{g1}(t_0,t)$ is analyzed in \emph{Case 1}. For $\bar V_{g2}(t_0,t)$, with   $\xi_i(j+1\leq i \leq \omega)$ having no upper bound, we obtain
\begin{equation}
\begin{aligned}
&V_g(t_0,t)\\
&=\sum\limits_{i = j+1}^\omega\int\limits_{t_0}^tg_i\mathcal{N}(\xi_i)\dot \xi_iE_{\alpha,\alpha}(-\lambda(t-\varsigma)^\alpha)(t-\varsigma)^{\alpha-1}d\varsigma\\
&\ \ \ \ -\sum\limits_{i = j+1}^\omega\int\limits_{t_0}^t\dot \xi_iE_{\alpha,\alpha}(-\lambda(t-\varsigma)^\alpha)(t-\varsigma)^{\alpha-1}d\varsigma.\\
\end{aligned}\label{69}
\end{equation}

We analyze the stability of $\bar V_{g2}(t_0,t)$ in \emph{Case 2}. Meanwhile, $\bar V_{g3}(t_0,t)$, which has the bounded variable $\delta_i(\omega+1\leq i \leq n)$ is denoted as
\begin{align}
&V_g(t_0,t)\nonumber\\
&=\sum\limits_{i = \omega+1}^n\int\limits_{t_0}^tg_i\mathcal{N}(\delta_i)\dot\delta_iE_{\alpha,\alpha}(-\lambda(t-\varsigma)^\alpha)(t-\varsigma)^{\alpha-1}d\varsigma\nonumber\\
&\ \ \ \ +\sum\limits_{i = \omega+1}^n\int\limits_{t_0}^t\dot\delta_iE_{\alpha,\alpha}(-\lambda(t-\varsigma)^\alpha)(t-\varsigma)^{\alpha-1}d\varsigma.\label{70}
\end{align}

We can know that  $\bar V_{g3}(t_0,t)$ is bounded from the results in \emph{Case 1}.

According to the analysis in \emph{Case 1} and the expression of $(80)$, we can also invoke a sequence to result in the contradiction. Therefore, it can be determined that $\delta_i(1\leq i \leq j)$ has an upper bound and $\delta_i(j+1\leq i \leq \omega)$ has a lower bound.

Combining the results in the three cases, we eventually conclude that $\delta_i$   must be bounded on $[t_0,t_f)$ for $i=1,2,\ldots,n$. Meanwhile, the boundedness of $V(t)$, $V_g(t)$ and $V_{gi}(t)$ can be obtained on $[t_0,t_f)$ for $i=1,2,\ldots,n$. Consequently, the proof is accomplished.



\ifCLASSOPTIONcaptionsoff
  \newpage
\fi


\begin{thebibliography}{10}

\bibitem{a0}
S. M. Mahmoud, \emph{Large-scale Control Systems: Theories and Techniques}. New York: Marcel Dekker, 1985.

\bibitem{a1}
M. Krstic, I. Kanellakopoulos and P. V. Kokotovi\'{c}, \emph{Nonlinear and Adaptive Control Design}. New York: Wiley, 1995.

\bibitem{a2}
C. Wen, ``Decentralized adaptive regulation," \emph{IEEE Trans. Autom. Control}, vol. 39, no. 10, pp. 2163-2166, Oct. 1994.


\bibitem{a6}
S. Jain and F. Khorrami, ``Decentralized adaptive output feedback design for large-scale nonlinear systems," \emph{IEEE Trans. Autom. Control}, vol. 42, no. 5, pp. 729-735, May 1997.

\bibitem{a5}
Z. P. Jiang, ``Decentralized and adaptive nonlinear tracking of large-scale systems via output feedback," \emph{IEEE Trans. Autom. Control}, vol. 45, no. 11, pp. 2122-2128, Nov. 2000.


\bibitem{a8}
J. Zhou and C. Wen, ``Decentralized backstepping sdaptive output tracking of interconnected nonlinear systems," \emph{IEEE Trans. Autom. Control}, vol. 53, no. 10, pp. 2378-2384, Nov. 2008.

\bibitem{interaction}
C. Wang and Y. Lin, ``Decentralized adaptive tracking control for a class of interconnected nonlinear time-varying systems," \emph{Automatica}, vol. 54, pp. 16-24, Apr. 2015.

\bibitem{aFu}
S. Tong, C. Liu and Y. Li, ``Fuzzy-Adaptive decentralized output-feedback control for large-scale nonlinear systems with dynamical uncertainties," \emph{IEEE Trans. Fuzzy Syst.}, vol. 18, no. 5, pp. 845-861, Oct. 2010.


\bibitem{aF1}
Z. Zhang, H. Liang, C. Wu, and C. K. Ahn, ``Adaptive event-triggered output feedback fuzzy control for nonlinear networked systems with packet dropouts and actuator failure," \emph{IEEE Trans. Fuzzy Syst.}, vol. 27, no. 9, pp. 1793-1806, Sep. 2019.


\bibitem{aF3}
S. Li, C. K. Ahn, and Z. Xiang, ``Sampled-data adaptive output feedback fuzzy stabilization for switched nonlinear systems with asynchronous switching," \emph{IEEE Trans. Fuzzy Syst.}, vol. 27, no. 1, pp. 200-205, Jan. 2019.


\bibitem{aF4}
Z. Fei, S. Shi, T. Wang, and C. K. Ahn, ``Improved stability criteria for discrete-time switched T-S fuzzy systems," IEEE Trans. Syst. Man Cybern. -Syst., to be published. doi:10.1109/TSMC.2018.2882630.

\bibitem{aF5}
S. Zheng, W. Li, ``Fuzzy finite time control for switched systems via adding a barrier power integrator," \emph{IEEE Trans. Cybern.}, vol. 49, no. 7, pp. 2693-2706, Jul. 2019.

\bibitem{aF2}
S. Zheng, P. Shi, S. Wang, and Y. Shi, ``Event triggered adaptive fuzzy consensus for interconnected switched multiagent systems," \emph{IEEE Trans. Fuzzy Syst.}, vol. 27, no. 1, pp. 144-158, Jan. 2019.



\bibitem{u1}
R. D. Nussbaum, ``Some remarks on a conjecture in parameter adaptive control," \emph{Syst. Control Lett.}, vol. 3, no. 5, pp. 243-246, Nov. 1983.

\bibitem{u2}
X. Ye and J. Jiang, ``Adaptive nonlinear design without a priori knowledge of control directions," \emph{IEEE Trans. Autom. Control}, vol. 43, no. 11, pp. 1617-1621, Nov. 1998.


\bibitem{u5}
X. Ye, ``Decentralized adaptive regulation with unknown high-frequency-gain signs," \emph{IEEE Trans. Autom. Control}, vol. 44, no. 11, pp. 2072-2076, Nov. 1999.


\bibitem{u4}
S. S. Ge and J. Wang, ``Robust adaptive tracking for time-varying uncertain nonlinear systems with unknown control coefficients," \emph{IEEE Trans. Autom. Control}, vol. 48, no. 8, pp. 1463-1469, Aug. 2003.




\bibitem{u9}
W. Chen, X. Li, W. Ren, and C. Wen, ``Adaptive consensus of multi-agent systems with unknown identical control directions based on a novel Nussbaum-type function," \emph{IEEE Trans. Autom. Control}, vol. 59, no. 7, pp. 1887-1892, Jul. 2014.

\bibitem{u10}
Z. Ding, ``Adaptive consensus output regulation of a class of nonlinear systems with unknown high-frequency gain," \emph{Automatica}, vol. 51, pp. 348-355, Jan. 2015.


\bibitem{u8}
C. Chen, Z. Liu, K. Xie, Y. Liu, Y. Zhang, and C. L. P. Chen, ``Adaptive fuzzy asymptotic control of MIMO systems with unknown input coefficients via a robust Nussbaum gain-based approach," \emph{IEEE Trans. Fuzzy Syst.}, vol. 25, no. 5, pp. 1252-1263, Oct. 2017.

\bibitem{u7}
J. Huang and Q. Wang, ``Decentralized adaptive control of interconnected nonlinear systems with unknown control directions," \emph{ISA Trans.}, vol. 74, pp. 60-66, Mar. 2018.

\bibitem{f1}
C. A. Monje, Y. Q. Chen, B. M. Vinagre, D. Y. Xue, and F. Vicente, \emph{Fractional-Order Systems and Controls: Fundamentals and Applications}. Springer, London, 2010.


\bibitem{fe2}
S. Zheng, ``{Robust stability of fractional order system with general interval uncertainties}," \emph{Syst. Control Lett.}, vol. 99, pp. 1-8, Jan. 2017.


\bibitem{f0}
I. Podlubny, \emph{Fractional Differential Equations}. San Diego, CA, USA: Academic Press, 1999.

\bibitem{fe1}
T. Lin and T. Lee, ``{Chaos synchronization of uncertain fractional-order chaotic systems with time delay based on adaptive fuzzy sliding mode control}," \emph{IEEE Trans. Fuzzy Syst.}, vol. 19, no. 4, pp. 623-635, Aug. 2011.


\bibitem{fe3}
Z. Yang, S. Zheng, F. Liu, Y. Xie, ``Adaptive output feedback control for fractional-order multi-agent systems," \emph{ISA Trans.}, to be published. doi:10.1016/J.ISATRA.2019.07.008.

\bibitem{fe4}
Z. Ma and H. Ma, "Reduced-order observer-based adaptive backstepping control for fractional-order uncertain nonlinear systems," \emph{IEEE Trans. Fuzzy Syst.}, to be published. doi:  10.1109/TFUZZ.2019.2949760.

\bibitem{f2}
M. \"{O}. Efe, ``{Fractional order systems in industrial automation―a survey}," \emph{IEEE Trans. Ind. Informat.}, vol. 7, no. 4, pp. 582-591, Nov. 2011.

\bibitem{f3}
D. Baleanu, J. A. T. Machado and A. C. J. Luo, \emph{Fractional Dynamics and Control}. New York, NY, USA: Springer, 2011.

\bibitem{f4}
D. Ding, D. Qi, Y. Meng, and L. Xu, ``{Adaptive Mittag--Leffler stabilization of commensurate fractional order nonlinear systems}," in \emph{Proc. IEEE 53rd Annu. Conf. Decis. Control (CDC)}, Los Angeles, CA, USA, 2014, pp. 6920-6926.

\bibitem{f7}
Y. Li, Y. Chen and I. Podlubny, ``Mittag--Leffler stability of fractional order nonlinear dynamic systems," \emph{Automatica}, vol. 45, no. 8, pp. 1965-1969, Aug. 2009.

\bibitem{f8}
Y. Li, Y. Chen and I. Podlubny, ``Stability of fractional order nonlinear dynamic systems: Lyapunov direct method and generalized Mittag--Leffler stability," \emph{Comput. Math. Appl.}, vol. 59, no. 5, pp. 1810-1821, Mar. 2010.

\bibitem{f9}
N. Aguila-Camacho, M. A. Duarte-Mermoud and J. A. Gallegos, ``Lyapunov functions for fractional order systems," \emph{Commun. Nonlinear Sci. Numer. Simul.}, vol. 19, no. 9, pp. 2951-2957, Sep. 2014.

\bibitem{f10}
H. Liu, Y. Pan, S. Li, and Y. Chen, ``Adaptive fuzzy backstepping control of fractional order nonlinear systems," \emph{IEEE Trans. Syst. Man Cybern. -Syst.}, vol. 47, no. 8, pp. 2209-2217, Aug. 2017.

\bibitem{f11}
Z. Ma and H. Ma, ``Adaptive fuzzy backstepping dynamic surface control of strict-feedback fractional order uncertain nonlinear systems," \emph{IEEE Trans. Fuzzy Syst.}, to be published. doi: 10.1109/TFUZZ.2019.2900602.

\bibitem{f12}
J. C. Trigeassou, N. Maamri, J. Sabatier, and A. Oustaloup, ``A Lyapunov approach to the stability of fractional differential equations," \emph{Signal Process.}, vol. 91, no. 3, pp. 437-445, Mar. 2011.

\bibitem{f5}
Y. Wei, Y. Chen, S. Liang, and Y. Wang, ``A novel algorithm on adaptive backstepping control of fractional order systems," \emph{Neurocomputing}, vol. 165, pp. 395-402, Oct. 2015.

\bibitem{f6}
Y. Wei, T. Peter, Z. Yao, and Y. Wang, ``Adaptive backstepping output feedback control for a class of nonlinear fractional order systems,"\emph{ Nonlinear Dyn.}, vol. 86, no. 2, pp. 1047-1056, Oct. 2016.

\bibitem{f17}
B. Liang, S. Zheng, Z. Yang, and F. Liu, ``Adaptive control for fractional-order interconnected systems," in \emph{Proc. 38th Chin. Control Conf.}, Guangzhou, China, 2019, pp. 2582-2587.



\bibitem{fa1}
M. \"{O}. Efe,  "Fractional fuzzy adaptive sliding-mode control of a 2-DOF direct-drive robot arm," \emph{IEEE Trans. Syst. Man Cybern. B (Cybern.)}, vol. 38, no. 6, pp. 1561-1570, Dec. 2008.





\bibitem{fa6}
N. Nikdel, M. Badamchizadeh, V. Azimirad, and M. A Nazari., ``Fractional-order adaptive backstepping control of robotic manipulators in the presence of model uncertainties and external disturbances," \emph{IEEE Trans. Ind. Electron.}, vol. 63, no. 10, pp. 6249-6256, Oct. 2016.


\bibitem{fa4}
B. Yang, T. Yu, H. Shu, D. Zhu, N. An, Y. Sang, and L. Jiang, ``Perturbation observer based fractional-order sliding-mode controller for MPPT of grid-connected PV inverters: Design and real-time implementation," \emph{Control Eng. Practice}, vol. 79, pp. 105-125, Oct. 2018.


\bibitem{motor_f}
P. Mani, R. Rajan, L. Shanmugam, and Y. H. Joo, ``Adaptive fractional fuzzy integral sliding mode control for PMSM model," \emph{IEEE Trans. Fuzzy Syst.}, vol. 27, no. 8, pp. 1674-1686, Aug. 2019.

\bibitem{motor_MIMO}
S. Cheng, Y. Wei, Y. Zhou, and Y. Wang, "Fractional order composite MRAC for MIMO systems based on SDU factorization," \emph{IFAC-PapersOnLine}, vol. 50, no. 1, pp. 7007-7012,  Jul. 2017.

\bibitem{Fuzzy}
L. X. Wang, ``Stable adaptive fuzzy control of nonlinear systems," \emph{IEEE Trans. Fuzzy Syst.}, vol. 1, no. 2, pp. 146-155, May 1993.

\bibitem{motor}
Z. Li, J. B. Park, Y. H. Joo, B. Zhang, and G. Chen, ``Bifurcations and chaos in a permanent-magnet synchronous motor," \emph{IEEE Trans. Circuits Syst. I: Fundam Theory Appl.}, vol. 49, no. 3, pp. 383-387, Mar. 2002.




\end{thebibliography}
\end{document}